\documentclass[12pt]{article}

\usepackage{amsfonts}
\usepackage{cite, natbib}
\usepackage[dvips]{graphicx}
\usepackage{keyval}
\usepackage{dsfont}
\usepackage{natbib}
\usepackage{multirow}
\usepackage{rotating}
\usepackage{endnotes}
\usepackage{amssymb,amsmath,lscape, bm, amsthm}
\usepackage{amsfonts}
\usepackage{graphicx}
\usepackage{color}
\usepackage{rotating}
\usepackage{multirow}
\usepackage{subfig}
\usepackage{caption}
\usepackage{setspace}
\usepackage{verbatim}
\usepackage{epstopdf}
\usepackage{booktabs}
\usepackage{array}
\usepackage[top=1in, bottom=1in, left=1.25in, right=1.25in]{geometry}
\usepackage{adjustbox}
\usepackage{dcolumn}
\usepackage{caption}
\usepackage{longtable}
\usepackage{mathrsfs}
\usepackage{pdfpages}

\newtheorem{assumption}{Assumption}[section]
\newtheorem{lemma}{Lemma}[section]

\newtheorem{algorithm}{Algorithm}[section]
\newtheorem{proposition}{Proposition}[section]
\newtheorem{remark}{Remark}[section]

\newcommand{\possessivecite}[1]{\citeauthor{#1}'s (\citeyear{#1})}

\begin{document}

\renewcommand{\baselinestretch}{1.5}
\normalsize

\renewcommand{\thefootnote}{\fnsymbol{footnote}}

\noindent{\Large{\sc{Regularizing stock return covariance matrices via multiple testing of correlations}} }

\bigskip
\bigskip

\renewcommand{\baselinestretch}{1.0}
\normalsize

\noindent{Richard Luger}\footnotemark[1]  

\smallskip

\noindent{Universit{\'e} Laval, Canada}

\noindent{}

\bigskip
\bigskip

\footnotetext[1]{Correspondence to: Department of Finance, Insurance and Real Estate, Laval University, Quebec City, Quebec G1V  0A6, Canada.

\emph{E-mail address:} {richard.luger@fsa.ulaval.ca}.

}

\bigskip
\bigskip

\renewcommand{\baselinestretch}{1.5}
\normalsize

\noindent{\bf Abstract:} 
This paper develops a large-scale inference approach for the regularization of stock return covariance matrices. 
The framework allows for the presence of heavy tails and multivariate GARCH-type effects of unknown form among the stock returns. 
The approach involves simultaneous testing of all pairwise correlations, followed by setting non-statistically significant elements to zero. 
This adaptive thresholding is achieved through sign-based  Monte Carlo resampling within multiple testing procedures, controlling either the traditional familywise error rate, a generalized familywise error rate, or the false discovery proportion. 
Subsequent shrinkage ensures that the final covariance matrix estimate is positive definite and well-conditioned while preserving the achieved sparsity. 
Compared to alternative estimators, this new regularization method demonstrates strong performance in simulation experiments and real portfolio optimization.

\bigskip
\bigskip


\bigskip


\bigskip
\bigskip
\bigskip
\bigskip

\renewcommand{\baselinestretch}{1.15}
\normalsize

{\footnotesize

\noindent{This paper is forthcoming in the Journal of Econometrics:}

\noindent{Luger, R. (2024). Regularizing stock return covariance matrices via multiple testing of correlations.
{\it Journal of Econometrics}, https://doi.org/10.1016/j.jeconom.2024.105753.
}
}

\thispagestyle{empty}

\newpage

\renewcommand{\baselinestretch}{1.75}
\normalsize

\pagenumbering{arabic}

\renewcommand{\thefootnote}{\arabic{footnote}}
\setcounter{footnote}{0}

\section{Introduction}

Estimating covariance matrices is a fundamental problem in multivariate statistical analysis, with wide-ranging applications in fields such as finance, economics, meteorology, climate research, spectroscopy, signal processing, pattern recognition, and genomics. In the realm of finance, accurate covariance matrix estimates are essential for capturing dependencies between asset returns -- a crucial input for portfolio optimization and risk management. Furthermore, many other statistical techniques rely on covariance matrix estimates, including regression analysis, discriminant analysis, principal component analysis, and canonical correlation analysis

Traditional estimation of the sample covariance matrix is known to perform poorly when the number of variables, $N$, is large compared to the number of observations, $T$. As the concentration ratio  $N/T$ grows, there are simply too many parameters relative to the available data points and the eigenstructure of the sample covariance matrix gets distorted in the sense that the sample eigenvalues are more spread out than the population ones \citep{Johnstone:2001}. The most egregious case occurs as $N/T > 1$, which causes the sample covariance matrix to become singular (non-invertible). By continuity, this matrix becomes ill-conditioned (i.e., its inverse incurs large estimation errors) as $N$ gets closer to $T$.

In such situations it is desirable to find alternative estimates that are more accurate and better conditioned than the sample covariance matrix.
Regularization methods for large covariance matrices can be divided into two broad categories: (i) methods that aim to improve efficiency and obtain well-conditioned matrices, and (ii) methods that introduce sparsity (off-diagonal zeros) by imposing special structures on the covariance matrix or its inverse (the precision matrix).
The first group includes linear shrinkage  \citep{Ledoit-Wolf:2003, Ledoit-Wolf:2004}, non-linear shrinkage \citep{Ledoit-Wolf:2012}, condition-number regularization  \citep{Won-Lim-Kim-Rajaratnam:2013}, and split-sample regularization \citep{Abadir-Distaso-Zikes:2014}. Methods that impose special structure include banding or tapering \citep{Bickel-Levina:2008b, Wu-Pourahmadi:2009} and thresholding \citep{Bickel-Levina:2008a, ElKaroui2008, Rothman-Levina-Zhu:2009, Cai-Liu:2011}, which  involves setting to zero the off-diagonal entries of the covariance matrix that are in absolute value below a certain data-dependent threshold.

\citet*{Bailey-Pesaran-Smith:2019}, hereafter BPS, develop an alternative thresholding approach using a multiple hypothesis testing procedure to 
assess the statistical significance of the elements of the sample correlation matrix; see also \citet[][p. 2748]{ElKaroui2008} who suggests a similar approach.
The idea is to test all pairwise correlations \emph{simultaneously}, and then to set to zero the elements that are not statistically significant. 
As with other thresholding methods, this multiple testing approach preserves the symmetry of the correlation matrix but it does not ensure its positive definiteness. BPS resolve this issue with an additional linear shrinkage step, whereby the correlation matrix estimator is shrunk towards the identity matrix to ensure positive definiteness. 
It must be emphasized that the BPS approach for reducing the number of spurious correlations is also of interest in the classical ``{low $N,$ large $T$}" setting.

The simultaneous testing of all pairwise correlations gives rise to a multiple comparisons problem. 
Indeed if the multiplicity of inferences is not taken into account, then the probability that some of the true null hypotheses (of zero pairwise correlation) are rejected by chance alone may be unduly large. A traditional objective in multiple testing is to control the familywise error rate (FWER), defined as the probability of rejecting at least one true null hypothesis. BPS use ideas from the multiple testing literature, but from the get-go they state in their introduction (p. 508) that they ``will not be particularly concerned with controlling the overall size of the joint $N(N-1)/2$ tests'' of zero pairwise correlations. The simulation evidence presented in this paper reveals that the empirical FWER with BPS thresholding can be severely inflated, resulting in far too many erroneous rejections of the null hypothesis of zero correlation. This over-rejection problem is greatly exacerbated by the presence of heavy tails, which obviously defeats the purpose of achieving sparsity.

Resampling techniques can be exploited to control the flood of {Type I} errors that arise when many hypothesis tests are performed simultaneously. 
In particular, such techniques can be used to account for the joint distribution of the test statistics and
obtain multiple testing procedures which achieve control of the  FWER and other false positive error rate measures; see \citet{Westfall-Young:1993},
\citet{Romano-Wolf:2005}, and \citet{Romano-Shaikh-Wolf:2008}. The primary goal of this paper is to extend the BPS  multiple testing regularization approach so that it is applicable to financial stock returns. Maintaining proper control of the FWER in this context means that more spurious correlations are detected and greater sparsity is induced.
More specifically, this paper makes two main contributions.

{\it First,} a sign-based  Monte Carlo resampling technique \citep{Dwass:1957,  Barnard:1963, Birnbaum:1974} is proposed to test pairwise correlations among stock returns.
The theory in BPS rules out the possibility of time-varying conditional variances and covariances -- a well-known feature of financial returns \citep{Cont:2001}. In turn, the presence of such effects gives rise to heavy tails and potential outliers in the distribution of returns. The  procedures in this paper are developed in a general framework that allows for the presence of heavy tails and multivariate GARCH-type effects of unknown form. Indeed, the Monte Carlo resampling scheme proceeds conditional on the absolute values of the centered returns, since only their signs are randomized. The \citet{Lehmann-Stein:1949} impossibility theorem shows that such sign-based tests are the \emph{only} ones that yield valid inference in the presence of non-normalities and heteroskedasticity of unknown form; see \citet{Dufour:2003} for more on this point.

{\it Second,} the resampling approach is used to develop both single-step and step-down test procedures that ensure control of two error rate measures advocated by \citet{Lehmann-Romano:2005}. The first of these measures is the $k$-FWER, defined as the probability of committing $k$ or more {Type I} errors, which are commonly referred to as false positives or false discoveries. 
Setting $k=1$ yields procedures controlling the traditional FWER used in confirmatory research, where the goal is to test a set of hypotheses while rigorously controlling the probability of making at least one false discovery. Simulation evidence reveals that the power of the step-down $1$-FWER test procedure is on par with that of the (FWER-adjusted) BPS tests. 
The second error rate measure is the \emph{false discovery proportion} (FDP), which represents the proportion of rejected null hypotheses that are erroneously rejected. Using the $k$-FWER (with $k > 1$) and the FDP as multiple testing criteria allows for a more lenient control over false rejections, thereby enhancing the ability to detect false null hypotheses. Unlike the traditional FWER, the $k$-FWER and the FDP offer a more nuanced approach to hypothesis testing in exploratory research. In these contexts, rejected hypotheses are generally not meant to be reported as end results, but are to be followed up with further validation experiments, such as  a subsequent out-of-sample performance evaluation; see \citet{Goeman-Solari:2011}.

Of course, the benefit of conducting multiple testing of correlations in terms of covariance matrix regularization is expected to increase with the true degree of sparsity in the population covariance matrix. In the context of stock portfolios, it is worth noting that this regularization approach provides a measure of diversification when returns tend to be positively correlated. Indeed the induced sparsity is expected to be proportional to the level of portfolio diversification, since in this case a well-diversified stock portfolio is precisely one in which the constituent assets demonstrate little or no correlation.

The rest of this paper is organized as follows. Section 2 presents the BPS multiple testing regularization approach. Section 3 establishes the  financial context and Section 4 develops the multiple testing procedures. Section 5 presents the results of simulation experiments that compare the performance of the new regularization method to BPS and other covariance matrix estimators. 
Section 6 further illustrates the large-scale inference approach with an out-of-sample evaluation of  portfolio optimization strategies. Section 7 offers some concluding remarks. 
All proofs, additional numerical results, and computing times for the multiple testing procedures are provided in the Supplementary material.

\section{Multiple testing regularization}

Consider a sample covariance matrix $\hat {\boldsymbol \Sigma} = [ \hat \sigma_{ij} ]_{N \times N}$  based a data sample of size $T$, and let  
$\hat {\boldsymbol \Gamma} =  [ \hat \rho_{ij} ]_{N \times N}$ denote the corresponding correlation matrix with typical element $ \hat \rho_{ij}= \hat{\sigma}_{ij}/\sqrt{\hat{\sigma}_{ii} \hat{\sigma}_{jj}  } $.  
As usual the sample covariance and correlation matrices are related via $\hat {\boldsymbol \Gamma} = \hat {\mathbf D}^{-1/2} \hat {\boldsymbol \Sigma}  \hat {\mathbf D}^{-1/2}$, where 
$\hat {\mathbf D} = \text{diag}(\hat \sigma^2_{1},\ldots,\hat \sigma^2_{N}  )$ with $\hat \sigma^2_{i}=\hat \sigma_{ii}$.
The BPS regularization strategy aims to improve $\hat {\boldsymbol \Sigma}  $ by  testing the family of $M= N(N-1)/2$ individual  hypotheses $  H_{ij}$ in the two-sided setting
\begin{equation}
 H_{ij}: \sigma_{ij} =0 \text{ versus }  H'_{ij}: \sigma_{ij}  \neq 0, \label{H0ij}
\end{equation}
for $j=1,\ldots,N-1$ and $i=j+1,\ldots,N$, while controlling the FWER. The elements that are found to be statistically insignificant are then set to zero. Instead of covariances, BPS prefer to base inference on the sample correlations since  they are all on the same scale. This leads to multiple testing procedures that are balanced in the sense that all constituent tests have about the same power \citep[][p. 50]{Westfall-Young:1993}. Note that the entries of the sample correlation matrix are intrinsically dependent even if the original observations are independent.

There are two types of FWER control. To introduce these, define an index $\ell$ taking values in the set $\mathcal M =\{1,\ldots, M \}$ as $j=1,\ldots,N-1$ and  $i=j+1,\ldots,N$  
so that $  H_{1} =  H_{2,1},  \ldots, H_{N-1} =  H_{N,1},H_{N} =  H_{3,2}, \ldots, H_{M} =  H_{N,N-1} $. Of these $M$ hypotheses tested, let $R$ denote the number of hypotheses rejected. Furthermore, let $\mathcal M_0 = \{\ell :  H_{\ell} \text{ is true}  \}$ denote the index set of true hypotheses. The number of false positive decisions (i.e., the number of {Type I} errors) is denoted by $F$. Given the nominal significance level $\alpha \in (0,1)$, the FWER is said to be controlled in the \emph{weak} sense when \sloppy 
$\Pr \left(  F \ge 1 \, \vert \, \bigcap_{\ell \in \mathcal M}  H_{\ell}  \right) = \Pr \left(   \text{Reject at least one }  H_{\ell} \, \vert \, \bigcap_{\ell \in \mathcal M}  H_{s}  \right) \le \alpha$, where the conditioning is on the \emph{complete} null hypothesis that $\mathcal M_0 =\mathcal M$ or $|\mathcal M_0| = M$.\footnote{The notation $\Pr(\text{event} \, \vert \, H)$ refers to the probability of the event occurring when $H$ holds true.} \emph{Strong} control is achieved when $\Pr \left(    F \ge 1 \, \vert \, \bigcap_{\ell \in \mathcal M_0}  H_{\ell}  \right) =   \Pr \left(   \text{Reject at least one }  H_{\ell}, \, \ell \in \mathcal M_0 \, \vert \, \bigcap_{\ell \in \mathcal M_0}  H_{\ell}  \right) \le \alpha$, regardless of the \emph{partial} null hypothesis (i.e., the particular intersection of hypotheses that happens to be true).

The BPS thresholding estimator, denoted here by $\hat {\boldsymbol \Gamma}_{\text{BPS}} $, has off-diagonal entries computed as
\begin{equation}
\hat \rho_{\text{BPS},ij} = \hat \rho_{ij} \mathds{1} \left\{ | \hat \rho_{ij} | > T^{-1/2} c_{\alpha}(N) \right\}, \label{bps}
\end{equation}
wherein $ \mathds{1}\{\cdot\}$ denotes the indicator function. 
The critical value $c_{\alpha}(N)$ appearing in (\ref{bps}) is given by
\begin{equation}
c_{\alpha}(N)  = \Phi^{-1}\left( 1 - \frac{\alpha}{2f(N)}  \right), \label{crit}
\end{equation}
 where  $\Phi^{-1}(\cdot)$ is the quantile function of a standard normal variate  and $f(N)$ is a general function of $N$ chosen to ensure  $\text{FWER} \le \alpha$ in the strong sense.
 Observe that the term  $T^{-1/2} c_{\alpha}(N)$ in (\ref{bps}) is a `universal'  threshold value in the sense that a single value is used to threshold 
all the off-diagonal elements of the correlation matrix $\hat {\boldsymbol \Gamma} $.
 Among other asymptotic properties, BPS  show that  $\hat {\boldsymbol \Gamma}_{\text{BPS}} $ converges to the true $ {\boldsymbol \Gamma} $ as the sample size grows.  As one  expects, the payoff in terms of noise reduction with this approach increases with the actual number of  zeros in  $ {\boldsymbol \Gamma} $.

When the number $M$ of tested hypotheses is very large and/or when strong evidence is not required, control of the FWER at conventional levels can be too stringent. 
In such cases one may wish to control the $k$-FWER, with $k$  set to a value greater than $1$ \citep{Lehmann-Romano:2005}.
Control of the $k$-FWER can also be either weak or strong. Given $\alpha \in (0,1)$ and $k \in \{1,\ldots, M\}$, weak control occurs whenever 
$\Pr \left(   F \ge k \, \vert \, \bigcap_{\ell \in \mathcal M}  H_{\ell}  \right) \le \alpha$ 
and strong control is achieved when 
$\Pr \left(   F \ge k \, \vert \, \bigcap_{\ell \in \mathcal M_0}  H_{\ell}  \right) \le \alpha$.

The choice of $k$ should be guided by the research objectives and the balance between minimizing the risk of false positives and maximizing the discovery of true effects. If the primary objective is to maintain strict control over {Type I} errors, then a constant, relatively low value of $k$ might be chosen regardless of the number of tested hypotheses. This ensures a consistent level of control over false positives. Conversely, when there is room for a controlled number of {Type I} errors and a desire to uncover more findings, a higher value of $k$ might be selected, for instance by letting $k$ increase with the number of tested hypotheses.

Another possibility considered in \citet{Lehmann-Romano:2005} is to maintain control of the FDP, defined as 
\begin{equation}
\text{FDP} = \left\{ \begin{array}{ll}
				{F}/{R}, & \text{ if } R > 0, \\
				          0, & \text{ if } R = 0.
				\end{array}
\right. \label{FDP}
\end{equation}
The FDP thus represents  the number of false rejections \emph{relative} to the total number of rejections (and equals zero if there are no rejections at all). 
FDP is a useful criterion when the focus is on controlling the rate at which null hypotheses are incorrectly rejected. 
Note that the FDP in (\ref{FDP})  is a random variable, and FDP control focuses on its tail probabilities.
For given values of  $\gamma \in [0,1)$  and $\alpha \in (0,1)$  specified by the user,
 strong control (of the tail probability) of the FDP means that
$\Pr \left(   \text{FDP} > \gamma  \, \vert \, \bigcap_{\ell \in \mathcal M_0}  H_{\ell}   \right) \le \alpha$. The interpretation is that, among the hypotheses that are rejected, the proportion of false discoveries may exceed $\gamma$ with probability no larger than $\alpha.$

Control of the FDP is a reasonable practice especially in non-confirmatory settings, where a certain proportion of false discoveries is considered acceptable.
The choice of $\gamma$, which determines the acceptable rate of false discoveries, involves a trade-off. A lower $\gamma$ reduces the risk of false discoveries but potentially missing some true effects. 
Indeed, setting $\gamma=0$ is  equivalent to  controlling the conservative $1$-FWER criterion. 
Conversely,  increasing the value of $\gamma$ allows for more discoveries but at the cost of accepting a higher rate of false discoveries.

\subsection{Positive definiteness}

As with other thresholding methods, the matrix  $\hat {\boldsymbol \Gamma}_{\text{BPS}}$ obtained via (\ref{bps}) is not necessarily 
well-conditioned nor even positive definite. 
 BPS solve this problem by shrinking $\hat {\boldsymbol \Gamma}_{\text{BPS}}$ towards $ \mathbf I_N$, the $N \times N$ identity matrix. 
Let $\lambda_{\min}(\hat {\boldsymbol \Gamma}_{\text{BPS}})$ denote the minimum eigenvalue of $\hat {\boldsymbol \Gamma}_{\text{BPS}}$
and set a limit $\epsilon > 0$ to avoid solutions that are too close to being singular.\footnote{In the simulation experiments and empirical application, the limit was set as $\epsilon = 0.01.$} The `shrinkage upon multiple testing' correlation matrix estimator is given by
\begin{equation}
\hat {\boldsymbol \Gamma}_\text{BPS}(\xi) = \xi \mathbf I_N + (1-\xi) \hat {\boldsymbol \Gamma}_{\text{BPS}}, \label{shrink}
\end{equation}
with shrinkage parameter $\xi \in [\xi_0,1]$, where $\xi_0 = \big(\epsilon- \lambda_{\min}(\hat {\boldsymbol \Gamma}_{\text{BPS}})\big)/\big(1-\lambda_{\min}(\hat {\boldsymbol \Gamma}_{\text{BPS}})\big)$ if $\lambda_{\min}(\hat {\boldsymbol \Gamma}_{\text{BPS}}) < \epsilon$, and  $\xi_0 = 0$ if 
$\lambda_{\min}(\hat {\boldsymbol \Gamma}_{\text{BPS}}) \ge \epsilon$. 
Note that by shrinking towards the  identity matrix, the resulting correlation matrix estimate preserves the  zeros (sparsity) achieved by $\hat {\boldsymbol \Gamma}_{\text{BPS}}$  and its diagonal elements do not deviate from unity. 
The computation of (\ref{shrink}) is made operational by replacing $\xi$ by $\xi^*$, which is found numerically as
\[
\xi^* = \arg \min_{\xi_0  \le \xi \le 1} \left\lVert  \hat{\boldsymbol \Gamma}^{-1}_0   - \hat {\boldsymbol \Gamma}_{\text{BPS}}^{-1}(\xi) \right\rVert^2_F,   
\]
where $\hat{\boldsymbol \Gamma}_0$ is a reference matrix and $\lVert \mathbf A \rVert_F$ denotes the Frobenius norm of  $\mathbf A $.\footnote{Recall that for a matrix $\mathbf A$  with 
elements $a_{ij}$, its  Frobenius norm is defined as $\lVert \mathbf A \rVert_F = \sqrt{  \sum_i \sum_j  a_{ij}^2  }$.
In the implementation, the value of $\xi^*$ was found by grid search with a step size of $\epsilon/2$.}

Following \citet{Schafer-Strimmer:2005}, the reference matrix $\hat{\boldsymbol \Gamma}_0$ is found by applying the linear shrinkage approach of \citet{Ledoit-Wolf:2004} to the sample correlation matrix, which yields
\[
\hat{\boldsymbol \Gamma}_0 = \hat \theta^* \mathbf I_N + (1-  \hat \theta^*) \hat {\boldsymbol \Gamma}, 
\]
where
\[
\hat \theta^* = 1- \frac{\underset{i \neq j}{\sum \sum} \hat \rho_{ij}  \left[  \hat \rho_{ij} - \frac{\hat \rho_{ij} (1-\hat \rho_{ij}^2 )}{2T}   \right] }{\frac{1}{T} \underset{i \neq j}{\sum \sum} (1- \hat \rho_{ij}^2 )^2 + \underset{i \neq j}{\sum \sum}\left[    \hat \rho_{ij} - \frac{\hat \rho_{ij} (1-\hat \rho_{ij}^2 )}{2T}  \right]^2},
\]
with the proviso that if $\hat \theta^* < 0$ then  $\hat \theta^*$ is set to $0$, and if $\hat \theta^* > 1$ then it is  set to $1$.\footnote{
The analytical expression for $\hat \theta^*$ is 
an estimate of the  optimal value of the shrinkage parameter that minimizes $\mathbb E (\lVert \hat{\boldsymbol \Gamma}_0 - {\boldsymbol \Gamma} \rVert^2_F)$, 
assuming  $\mathbb E(\hat \rho_{ij})$ can be approximated by $\hat \rho_{ij} - \hat \rho_{ij}(1-\hat \rho^2_{ij})/2T$ and 
$\text{Var}(\hat \rho_{ij})$  by $(1-\hat \rho_{ij}^2)^2/T$ \citep[cf.][]{Soper-Young-Cave-Lee-Pearson:1917}.
As \citet{Ledoit-Wolf:2003} explain, most shrinkage covariance matrix estimators  are based on such first-cut assumptions.
}
The resulting covariance matrix estimate is given by $\hat {\boldsymbol \Sigma}_{\text{BPS}}(\xi^*) =      \hat {\mathbf D}^{1/2}  \hat {\boldsymbol \Gamma}_{\text{BPS}}(\xi^*)  \hat {\mathbf D}^{1/2},$ wherein $ \hat {\boldsymbol \Gamma}_{\text{BPS}}(\xi^*)$ corresponds to (\ref{shrink}) evaluated with $\xi^*$.

\section{Financial context}

Consider $N$ financial assets  with time-$t$  returns   $\mathbf r_t =(r_{1,t},\ldots,r_{N,t})^{\prime}$, for $t=1,\ldots,T$, decomposed as
\begin{equation}
\begin{split}
\mathbf r_t  & = \boldsymbol \mu    + \boldsymbol \varepsilon_t, \\
 \boldsymbol \varepsilon_t & = \boldsymbol \Sigma_t^{1/2} \mathbf z_t,
\end{split}
\label{model}
\end{equation}
where $\boldsymbol \mu  = \mathbb E(\mathbf r_t) = \big(\mu_{1}, \ldots, \mu_{N}\big)^{\prime}$ is a vector of location parameters.
The error $ \boldsymbol \varepsilon_t = (\varepsilon_{1,t},\ldots,\varepsilon_{N,t})^{\prime}$ in (\ref{model}) consists of an innovation vector $\mathbf z_t$  satisfying Assumption \ref{assumption1}  below, and an unspecified $N \times N$ ``square root" matrix $\boldsymbol \Sigma_t^{1/2} $ such that 
 $\boldsymbol \Sigma_t^{1/2} (\boldsymbol \Sigma_t^{1/2})^{\prime} = \boldsymbol \Sigma_t = \mathbb E(\boldsymbol  \varepsilon_t \boldsymbol  \varepsilon_t^{\prime} \, \vert \, \mathcal F_{t-1})$, where $\mathcal F_{t-1} = (\mathbf r_{t-1}^{\prime},\mathbf r_{t-2}^{\prime}, \ldots)^{\prime}$.
This framework is compatible with several popular models of time-varying covariances, such as multivariate GARCH models \citep{Silvennoinen-Terasvirta:2009,  Boudt-Galanos-Payseur-Zivot:2019} and multivariate stochastic volatility models  \citep{Chib-Omori-Asai:2009}.

\begin{assumption}
\label{assumption1}
The innovations $ \{ \mathbf z_t \}_{t=1}^T$ are independently (but not necessarily identically) distributed according to spherically symmetric distributions, with moments
 $\mathbb E(\mathbf  z_t \, \vert \,  \mathcal F_{t-1}) = \boldsymbol 0$ and $\mathbb E(\mathbf  z_t \mathbf  z_t^{\prime} \, \vert \,  \mathcal F_{t-1}) = \mathbf I_N$, 
 for $t=1,\ldots,T$.
\end{assumption}

\noindent This assumption means that $\mathbf  z_t$ admits the stochastic representation $\mathbf  z_t \stackrel{d}{=} \mathbf  H  \mathbf  z_t $, 
where the symbol $\stackrel{d}{=}$ stands for an equality in distribution and $\mathbf  H $ is any $N \times N$ orthogonal matrix such that 
$\mathbf  H^{\prime} \mathbf  H = \mathbf  H \mathbf  H^{\prime} =  \mathbf I_N$. This class includes the   multivariate versions of the standardized 
normal, Student $t$, logistic, and {Laplace} distributions, among many others \citep{Fang-Kotz-Ng:1990}.

When  $\mathbf  z_t$ has a well-defined density, then Assumption \ref{assumption1} is equivalent to assuming that the conditional distribution of  $\mathbf r_t  $ is elliptically symmetric, meaning that its density has the form
$|\boldsymbol \Sigma_t^{-1/2}| f\big( (\mathbf r_t  - \boldsymbol \mu   )^{\prime}  \boldsymbol \Sigma_t^{-1} (\mathbf r_t  - \boldsymbol \mu   ) \big)$ for some 
non-negative scalar function $f(\cdot)$.
Elliptically symmetric distributions play a very important role in mean-variance analysis (cf. Section 6) because they guarantee full compatibility with expected utility maximization regardless of investor preferences
\citep{Berk:1997, Chamberlain:1983, Owen-Rabinovitch:1983}.

In the context of (\ref{model}), the complete null hypothesis is formally stated as
\begin{equation}
H_0: \boldsymbol \Sigma_t^{1/2} = \mathbf D_t^{1/2},  \label{null}
\end{equation}
for $t=1,\ldots,T$, where $\mathbf D_t^{1/2}$ is a diagonal matrix (i.e., with zeros outside the main diagonal).
Observe that conditional heteroskedasticity is permitted under $H_0$; i.e., the diagonal elements of $\mathbf D_t^{1/2}$ may be time-varying. 
It is easy to see that when Assumption \ref{assumption1}  holds and $H_0$ is true,   the error vector  $\boldsymbol \varepsilon_t =(\varepsilon_{1,t},\ldots,\varepsilon_{N,t})^{\prime}$ becomes sign-symmetric \citep{Serfling:2006} in the sense that
\[
\boldsymbol \varepsilon_t  \stackrel{d}{=} \mathbb S_t \boldsymbol \varepsilon_t, 
\]
for $t=1,\ldots,T$ and for all $N \times N$ diagonal matrices  $ \mathbb S_t $  with $\pm 1$ on the diagonal.

\begin{assumption}
\label{assumption2}
The unconditional covariance matrix $\boldsymbol \Sigma= \mathbb E\left( \boldsymbol \varepsilon_t \boldsymbol \varepsilon_t^{\prime} \right)$ exists.
\end{assumption}

\noindent With this assumption the sign-symmetry condition 
$(\varepsilon_{i,t}, \varepsilon_{j,t})  \stackrel{d}{=} (\pm \varepsilon_{i,t}, \pm \varepsilon_{j,t}) $ implies  $\sigma_{ij}=0$, for $i \neq j$, where  $\sigma_{ij}$ is the 
$(i,j)^\text{th}$ element of $\boldsymbol \Sigma$ \citep[][Lemma 1.3.28]{Randles-Wolfe:1979}.

\section{Multiple testing procedures}

Assume momentarily that the true value ${\boldsymbol \mu}^{\ast}$ of  $\boldsymbol \mu$ in (\ref{model}) is known.
For instance, with daily returns it is often reasonable to assume that ${\boldsymbol \mu}^{\ast} = \boldsymbol 0$. 
The case of unknown location parameters will be dealt with in Section 4.6.

Given the value of ${\boldsymbol \mu}^{\ast}$, centered returns can then be defined 
as $\mathbf y_t = \mathbf r_t  - {\boldsymbol \mu}^{\ast}  = (y_{1,t},\ldots,y_{N,t})^{\prime}$, for $t=1,\ldots,T$, and these have the same properties as 
$ \boldsymbol \varepsilon_t$.
The time series of centered returns are collected into the $T \times N$ matrix $\mathbf Y = [\mathbf y_1,\ldots,\mathbf y_T  ]^{\prime}$.
Following BPS, inference  is based on the pairwise correlations
$
\hat \rho_{ij} = { \hat \sigma_{ij} }/{\sqrt{\hat \sigma_{ii}  \hat \sigma_{jj}}  }
$
that constitute the  matrix $\hat {\boldsymbol \Gamma} =  [ \hat \rho_{ij} ]_{N \times N}$. 
This matrix can be obtained from the familiar relationship
$\hat {\boldsymbol \Gamma} = \hat {\mathbf D}^{-1/2} \hat {\boldsymbol \Sigma}  \hat {\mathbf D}^{-1/2}$ with 
$\hat {\mathbf D} = \text{diag}(\hat \sigma^2_{1},\ldots,\hat \sigma^2_{N}  )$ and $\hat \sigma^2_{i}=\hat \sigma_{ii}$, where $\hat {\boldsymbol \Sigma}$ now comprises the variances and covariances about the origin, computed as $ \hat \sigma_{ij} = T^{-1} \sum_{t=1}^T y_{i,t} y_{j,t}$, for $i,j=1,\ldots,N$.

Let $\tilde{ \mathbb S}_t = \text{diag}(\tilde s_{1,t},\ldots,\tilde s_{N,t})$, for $t=1,\ldots,T$, where
$\tilde s_{i,t} $ are independent Rademacher random draws such that $\Pr(\tilde s_{i,t} =1) = \Pr(\tilde s_{i,t} =-1 ) =1/2$, for each $i, t$. An artificial sample $\tilde{\mathbf Y}  = \left[\tilde{\mathbf y}_1,\ldots,\tilde{\mathbf y}_T  \right]^{\prime}$ with $\tilde { \mathbf y}_t  = ( \tilde y_{1,t},\ldots, \tilde y_{N,t})'$
is then defined as
\begin{equation}
\tilde{\mathbf Y}  = \left[ \tilde{ \mathbb S}_1 {\mathbf y}_1,\ldots, \tilde{ \mathbb S}_T {\mathbf y}_T  \right]^{\prime}.  \label{ytil}
\end{equation}

If  Assumption \ref{assumption1} holds and $H_0$ is true, then  $\mathbf Y \stackrel{d}{=} \tilde{\mathbf Y} $, for each of the $2^{TN}$ possible matrix realizations of $ \tilde{\mathbf Y} $, given  $|\mathbf Y |$.
Here $|\mathbf Y |$ is the matrix of entrywise absolute values of $\mathbf Y$.
For a given artificial sample $\tilde { \mathbf Y}$, let 
$\tilde {\boldsymbol \Gamma} =  [ \tilde \rho_{ij} ]_{N \times N}$ denote the associated correlation matrix comprising the pairwise correlations about the origin
$\tilde \rho_{ij} = { \tilde \sigma_{ij} }/{\sqrt{\tilde \sigma_{ii}  \tilde \sigma_{jj}}  }$,  where $\tilde \sigma_{ij} = {T}^{-1} \sum_{t=1}^T \tilde y_{i,t}  \tilde y_{j,t}$.

\begin{proposition}
\label{proposition1}
Suppose that (\ref{model}) holds along with Assumption \ref{assumption1}, and  
consider  $\tilde { \mathbf Y}$ generated according to (\ref{ytil}). 
If  $H_0$ in (\ref{null}) is true, then 
 $ \Pr(\hat {\boldsymbol \Gamma} = \tilde {\boldsymbol \Gamma} \, \vert \, |\mathbf Y|  )= 1/2^{TN} $.
Furthermore under Assumption \ref{assumption2},  $\mathbb E( \tilde y_{i,t} \tilde y_{j,t} \,  \vert \,  |\mathbf Y| ) = 0$, for $i \neq j$.
\end{proposition}

Proposition \ref{proposition1} shows that $\hat {\boldsymbol \Gamma} $ is conditionally \emph{pivotal} under $H_0$, meaning that its sign-randomization distribution does not depend on any nuisance parameters. In principle, critical values could be found from the conditional distribution of  $\hat {\boldsymbol \Gamma} $ derived from the $2^{TN}$  equally likely values represented by $ \tilde {\boldsymbol \Gamma} $. Determination of this distribution from a complete enumeration of all possible realizations of  $\tilde {\boldsymbol \Gamma} $  is obviously impractical. 
Following \citet{Zhu-Neuhaus:2000}, the algorithms developed next use a non-parametric  Monte Carlo  test technique in order to circumvent this problem and still obtain exact $p$-values.

\subsection{Unadjusted $p$-values}

It is useful to first describe how to obtain the Monte Carlo $p$-values without multiplicity adjustments,  even if they are not used directly for multiple testing regularization. Indeed, these raw  $p$-values are the foundational blocks for the  development of multiplicity-adjusted $p$-values.
Some additional notation will facilitate the explanation of the Monte Carlo test technique.
With a correlation matrix ${\boldsymbol \Gamma} = [\rho_{ij}]_{N \times N}$ as input, let $\boldsymbol \rho = \text{vechs}({\boldsymbol \Gamma} ) = (\rho_{2,1},\ldots,\rho_{N,1}, \rho_{3,2}, \ldots, \rho_{N,N-1})^{\prime}$ be the $M \times 1$ vector resulting from its strict half-vectorization and let $\text{fill}(\cdot)$ denote the inverse function such that 
${\boldsymbol \Gamma} = \text{fill}(\text{vechs}({\boldsymbol \Gamma} ))$.
Furthermore, let $ (\hat \rho_{1},\ldots, \hat \rho_{M})^{\prime} = \text{vechs}(\hat {\boldsymbol \Gamma} )$ so that $|\hat \rho_{\ell}|$ is the statistic for $H_{\ell}$,  $\ell=1,\ldots,M$.

Note that  sampling according to (\ref{ytil}) yields a discrete distribution of $\tilde {\boldsymbol \Gamma}$ values, which means that ties among the resampled values can occur, at least theoretically. Following \citet{Dufour:2006}, these are dealt with by working with lexicographic (tie-breaking) ranks.
Algorithm \ref{algorithm1} details the steps to obtain the unadjusted Monte Carlo $p$-values with
$B-1$ resampling draws chosen so that  $\alpha B$ is an integer, where $\alpha \in (0,1)$ is the desired significance level.

\begin{algorithm}[Unadjusted Monte Carlo $p$-values]
\label{algorithm1}
\leavevmode 
\begin{enumerate}
{\normalfont

\item For $b=1,\ldots,B-1$, repeat the following steps:

  \begin{itemize}
 
 \item[(a)] generate an artificial data sample $\tilde{\mathbf Y}_b$  according to (\ref{ytil}); 
 
 \item[(b)] compute the associated  matrix of correlations about the origin $\tilde {\boldsymbol \Gamma}_b =  [ \tilde \rho_{ij,b} ]_{N \times N}$ and let 
 $ (\tilde \rho_{1,b},\ldots, \tilde \rho_{M,b})^{\prime} = \text{vechs}(\tilde {\boldsymbol \Gamma}_b )$.
  
 \end{itemize}

\item Draw $u_b \sim \mathcal U(0,1),$ for $b=1,\ldots,B.$

\item For  $\ell=1,\ldots,M$,

  \begin{itemize}
 
 \item[(a)] create the pairs $(|\tilde \rho_{\ell,1}|, u_1 ),\ldots,(|\tilde \rho_{\ell,B-1}|, u_{B-1} ), (|\hat \rho_{\ell}|, u_B )$, and
 compute the lexicographic rank of $| \hat \rho_{\ell} | $ among the $| \tilde \rho_{\ell} | $'s as
 \[
  R_\text{U}( | \hat \rho_{\ell} | ) = 1 + \sum_{b=1}^{B-1} \mathds{1} \left \{  | \hat \rho_{\ell} |  >  | \tilde \rho_{\ell,b} |   \right\}  
 + \sum_{b=1}^{B-1} \mathds{1} \left\{  | \hat \rho_{\ell} |  =  | \tilde \rho_{\ell,b} |   \right\}   \mathds{1} \left\{ u_B > u_b  \right\};
 \]

 \item[(b)] compute the unadjusted Monte Carlo $p$-value of $|\hat \rho_{\ell}|$ as
  \[
  \tilde p_\text{U}( | \hat \rho_{\ell} |)  = \frac{B-    R_\text{U}( | \hat \rho_{\ell} |) + 1}{B}. 
  \]
  
 \end{itemize}

}
\end{enumerate}
\end{algorithm}

\begin{remark} 
\normalfont
As the next proposition shows, setting $B=20$ is sufficient to obtain a test with exact level 0.05. A larger number of replications decreases the test's sensitivity  to the underlying randomization and typically leads to power gains.  Unreported simulation results reveal that increasing $B$ beyond 100 has only a small effect on power.
\end{remark}

\begin{proposition}
\label{proposition2}
Suppose that (\ref{model}) holds along with Assumptions \ref{assumption1} and \ref{assumption2}. Then the Monte Carlo $p$-values computed according to Algorithm \ref{algorithm1} guarantee that 
 $\Pr(   \tilde p_\text{U}(| \hat \rho_{\ell} | )  \le \alpha   \, \vert \, H_0  ) = \alpha$,  for  $\ell=1,\ldots,M$.
\end{proposition}

The  Monte Carlo $p$-values obtained from Algorithm \ref{algorithm1} have the usual interpretation: $  \tilde p_\text{U}(| \hat \rho_{\ell} | )$ is the proportion of $| \tilde \rho_{\ell} |$ values as extreme or more extreme than the observed $| \hat \rho_{\ell} |$ value in its resampling distribution.

\subsection{Single-step adjusted $p$-values for $k$-FWER control}

\citet{Westfall-Young:1993} propose several resampling-based methods to adjust $p$-values so as to account for multiplicity. 
Adjusted $p$-values are defined as the smallest significance level for which one still rejects an individual hypothesis $H_{\ell}$, given a particular multiple test procedure. 
Let $\text{$k$-max}(|\boldsymbol \rho|)$ denote the $k^\text{th}$ largest value of $|\boldsymbol \rho|$. 
So if the elements of $|\boldsymbol \rho|$ are sorted in decreasing order as $|\rho_{(1)}| \ge |\rho_{(2)}| \ge \ldots \ge |\rho_{(M)}|$, then $\text{$k$-max}(|\boldsymbol \rho|)=|\rho_{(k)}|$.

A straightforward extension of Westfall and Young's \emph{single-step} (SS) \emph{maxT adjusted p-values} to the present context for $k$-FWER control 
 yields the definition 
\begin{equation}
 p^k_\text{SS}( | \hat \rho_{\ell} | ) = \Pr \left( \text{$k$-max}(|\tilde{ \boldsymbol \rho}|)     \ge  |\hat \rho_{\ell}|  \, \big\vert \, H_0  \right), \label{pSS}
\end{equation}
for $\ell=1,\ldots,M$, where $\tilde{\boldsymbol \rho} = \text{vechs}(\tilde {\boldsymbol \Gamma} )$ and
$H_0$ is the complete null hypothesis in (\ref{null}).
When $k=1$, (\ref{pSS}) reduces to Definition (2.8) in \citet{Westfall-Young:1993}. 
In words, this says that the SS adjusted $p$-value is the probability that the 
$k^\text{th}$ largest absolute correlation in the artificial data is greater than the observed absolute correlation in the actual data.

\begin{lemma}
\label{lemma1}
Let $\hat m^k = \text{$k$-max}(|\hat{ \boldsymbol \rho}|)  $ with $ \hat{\boldsymbol \rho} = \text{vechs}(\hat {\boldsymbol \Gamma} )$ and
suppose that (\ref{model}) holds along with Assumptions \ref{assumption1} and \ref{assumption2}. 
Then, under  (\ref{null}), 
$\hat m^k \stackrel{d}{=} \tilde m^k$, where 
$\tilde m^k =\text{$k$-max}(|\tilde{ \boldsymbol \rho}|)  $,  $ \tilde{\boldsymbol \rho} = \text{vechs}(\tilde {\boldsymbol \Gamma} )$, and
 $\tilde {\boldsymbol \Gamma}$ is computed from an artificial sample $\tilde{\mathbf Y}$  generated according to (\ref{ytil}).
Moreover, upon application of Algorithm \ref{algorithm1}, it follows that $ \Pr \left( \tilde p_\text{U}\left( \hat m^k \right)  \le \alpha  \, \vert \, H_0  \right) = \alpha$.
\end{lemma}

Lemma \ref{lemma1} paves the way for  the computation of the Monte Carlo version of (\ref{pSS}) as described next.
The number of resampling draws $B-1$ is assumed to be chosen so that $\alpha B$ is an integer, where $\alpha \in (0,1)$ is the desired $k$-FWER.

\begin{algorithm}[Single-step $k$-FWER-adjusted Monte Carlo $p$-values]
\label{algorithm2}
\leavevmode 
\begin{enumerate}
{\normalfont

\item For $b=1,\ldots,B-1$, repeat the following steps:
  \begin{itemize}
 
 \item[(a)] generate an artificial data sample $\tilde{\mathbf Y}_b$  according to (\ref{ytil});
 
 \item[(b)] compute the associated  matrix of correlations about the origin $\tilde {\boldsymbol \Gamma}_b =  [ \tilde \rho_{ij,b} ]_{N \times N}$ and let 
 $ \tilde{\boldsymbol \rho}_b = \text{vechs}(\tilde {\boldsymbol \Gamma}_b )$;
 
 \item[(c)] find $\tilde m^k_{b} = \text{$k$-max}(|\tilde{ \boldsymbol \rho}_b|)$.
  
 \end{itemize}
 
\item Draw $u_b \sim \mathcal U(0,1),$ for $b=1,\ldots,B.$

\item For  $\ell=1,\ldots,M$,

\begin{itemize}

\item[(a)] create the pairs $(\tilde m^k_{1}, u_1 ),\ldots,(\tilde m^k_{B-1}, u_{B-1} ), (|\hat \rho_{\ell}|, u_B )$, and compute   the lexicographic rank of $| \hat \rho_{\ell} | $ among the $\tilde m^k_b$'s as
 \[
 R^k_\text{SS}( | \hat \rho_{\ell} | ) = 1 + \sum_{b=1}^{B-1} \mathds{1} \left\{  | \hat \rho_{\ell} |  >  \tilde m^k_{b}   \right\}  + \sum_{b=1}^{B-1} \mathds{1} 
 \left\{  | \hat \rho_{\ell} |  =  \tilde m^k_{b}  \right\}   \mathds{1} \left\{ u_B > u_b  \right\};
\]

\item[(b)] compute the SS adjusted Monte Carlo $p$-value of $|\hat \rho_{\ell}|$  as
  \[
  \tilde p^k_\text{SS}(| \hat \rho_{\ell} | )  = \frac{B -  R^k_\text{SS}( | \hat \rho_{\ell} |) + 1}{B}.
  \]

\end{itemize}
}
\end{enumerate}
\end{algorithm}

\begin{remark}
\normalfont
Finding the {$k$-max} in Step 1-(c) can be done simply by sorting the elements of $|\tilde{ \boldsymbol \rho}_b|$ in decreasing order and then outputting the $k^{\text{th}}$ element in the sorted array. This selection problem can be solved in $O(M \log M)$ average time by sorting the $M$ values using a {Quicksort} or {Heapsort} algorithm. Alternatively, one can use the 
 {Quickselect} algorithm, which has an average-case time complexity of $O(M)$; see \citet[][Ch. 9]{CLRS:2022}, among others.  
\end{remark}

\begin{proposition}
\label{proposition3}
Under (\ref{model}) and Assumptions \ref{assumption1} and \ref{assumption2}, Algorithm \ref{algorithm2} has weak control of the finite-sample $k$-FWER in the sense that 
$\Pr \left(   \text{Reject at least $k$ hypotheses } H_{\ell}  \, \vert \, H_{0} \right) = \alpha$.
\end{proposition}

The proof in \citet[][p. 53]{Westfall-Young:1993}  that their SS adjusted $p$-values control the FWER in the strong sense relies heavily on 
the assumption of \emph{subset pivotality}. That is, they assume that the joint distribution of unadjusted $p$-values under any partial null hypothesis is identical to that under the complete null hypothesis. As noted by \citet[][p. 43, Example 2.2]{Westfall-Young:1993} and  \citet[][Example 7]{Romano-Wolf:2005}, this assumption fails in the context of testing pairwise correlations. However, as the next proposition shows, subset pivotality is not a necessary condition for strong control; see also \citet{Romano-Wolf:2005}, \citet{Westfall-Troendle:2008}, and \citet{Goeman-Solari:2010}.

\begin{proposition}
\label{proposition4}
Under (\ref{model}) and Assumptions \ref{assumption1} and \ref{assumption2},  
Algorithm \ref{algorithm2} maintains strong control of the finite-sample $k$-FWER; i.e., 
$\Pr \left(   \text{Reject at least $k$ hypotheses } H_{\ell}, \, \ell \in \mathcal M_0 \, \vert \, \bigcap_{\ell \in \mathcal M_0}  H_{\ell}  \right) 
\le \alpha$.
\end{proposition}

The proof of this result (given in the  Supplementary material) makes clear that the SS procedure becomes more conservative as the cardinality of $\mathcal M_0$ decreases.

\subsection{Step-down adjusted $p$-values for $k$-FWER control}

A disconcerting feature of (\ref{pSS}) is that all the $p$-values are adjusted according to the distribution of the $k^\text{th}$ largest absolute correlation. 
Potentially less conservative $p$-values may be obtained from step-down adjustments that result in  uniformly smaller $p$-values, while retaining the same protection against {Type I} errors. 
The underlying idea is analogous to \possessivecite{Holm:1979} sequential refinement of the Bonferroni adjustment, which eliminates from consideration any null hypotheses that are rejected at a previous step.\footnote{\citet{Romano-Wolf:2005} discuss the \emph{idealized} step-down method; see also \citet{Romano-Wolf:2016}. That method is not feasible with the  resampling scheme developed here, because it is not possible to generate artificial data that obey the null hypothesis for each possible intersection 
of true null hypotheses.}

With the absolute correlations $|\hat \rho_{1}|,\ldots,|\hat \rho_{M}|$,  let the ordered test statistics have index values $\pi_1,\ldots,\pi_M$ so that $|\hat \rho_{\pi_1}| \ge |\hat \rho_{\pi_2}| \ge \ldots \ge |\hat \rho_{\pi_M}|$.
To control the $k$-FWER, the definition of  Westfall and Young's \emph{step-down} (SD) \emph{maxT adjusted p-values} can be 
extended as follows:
\begin{equation}
\begin{split}
p^k_{\text{SD}}(|\hat \rho_{\pi_1}|) & = \Pr \bigl( \tilde m^k_1 \ge |\hat \rho_{\pi_1}| \, \vert \, H_0   \bigr), \text{ with } \tilde m^k_1 =\text{$k$-max}\bigl(|\tilde{ \boldsymbol \rho}|\bigr),  \\
p^k_{\text{SD}}(|\hat \rho_{\pi_2}|) & = \Pr \bigl( \tilde m^k_2 \ge |\hat \rho_{\pi_2}| \, \vert \, H_0   \bigr), \text{ with } \tilde m^k_2 =\tilde m^k_1,  \\
& \vdots \\
p^k_{\text{SD}}(|\hat \rho_{\pi_k}|) & = \Pr \bigl( \tilde m^k_k \ge |\hat \rho_{\pi_k}| \, \vert \, H_0   \bigr), \text{ with } \tilde m^k_k =\tilde m^k_1,  \\
p^k_{\text{SD}}(|\hat \rho_{\pi_{k+1}}|) & = \max \Bigl[  p^k_{\text{SD}}\bigl(|\hat \rho_{\pi_k}|\bigr),  \, \Pr \bigl(  \tilde m^k_{k+1} \ge |\hat \rho_{\pi_{k+1}}| \, \vert \, H_0  \bigr)    \Bigr], \\
										 & \quad \quad\quad\quad\quad\quad\quad\quad\quad\quad\quad \quad\quad\quad\quad\quad\quad \text{ with }  \tilde m^k_{k+1}=\min\bigl( \tilde m^k_k, \max_{s=k+1,\ldots,M} |\tilde \rho_{\pi_s}| \bigr), \\
					& \vdots \\
p^k_{\text{SD}}(|\hat \rho_{\pi_\ell}|) & = \max \Bigl[  p^k_{\text{SD}}\bigl(|\hat \rho_{\pi_{\ell-1}}|\bigr),  \, \Pr \bigl(  \tilde m^k_{\ell} \ge |\hat \rho_{\pi_\ell}| \, \vert \, H_0  \bigr)    \Bigr], \\
                                        & \quad \quad \quad\quad\quad\quad\quad\quad\quad\quad\quad\quad \quad\quad\quad\quad\quad\quad  \text{ with }  \tilde m^k_{\ell}=\min\bigl( \tilde m^k_{\ell-1}, \max_{s=\ell,\ldots,M} |\tilde \rho_{\pi_s}| \bigr),  \\
                    & \vdots \\	
p^k_{\text{SD}}(|\hat \rho_{\pi_M}|) & = \max \Bigl[  p^k_{\text{SD}}\bigl(|\hat \rho_{\pi_{M-1}}|\bigr),  \, \Pr \bigl(  \tilde m^k_{M} \ge |\hat \rho_{\pi_M}| \, \vert \, H_0  \bigr)    \Bigr], \text{ with }  \tilde m^k_{M}=\min\bigl( \tilde m^k_{M-1},  |\tilde \rho_{\pi_M}| \bigr), 
\end{split} \label{pSD}
\end{equation}
wherein the sequence of index values $\pi_1,\ldots,\pi_M$ is  held \emph{fixed}.\footnote{That is, the adjustments are made by ``stepping down'' from the largest test statistic to the smallest.} 

Instead of adjusting all $p$-values according to the distribution of the $k^\text{th}$ largest  absolute correlation, this approach only adjusts the $p$-values for $|\hat \rho_{\pi_1}|, |\hat \rho_{\pi_2}|, \ldots, |\hat \rho_{\pi_k}|$ using this distribution. 
The remaining $p$-values in steps $k+1,\ldots, M$  are then adjusted according to the distributions of smaller and smaller sets of maximum absolute correlations, 
where the $\min$ operator is used to ensure that $ \tilde m^k_{\ell}  \le \tilde m^k_{\ell-1}  \le  \text{$k$-max}\bigl(|\tilde{ \boldsymbol \rho}|\bigr),$ for $\ell=k+1,\ldots, M.$

Note that the SD adjusted $p$-values have the same step-down monotonicity as the original test statistics; i.e., smaller $p$-values are associated with larger values of the  $|\hat \rho_{\ell}|$ test statistics. This is obvious for  $p^k_{\text{SD}}(|\hat \rho_{\pi_1}|), \ldots, p^k_{\text{SD}}(|\hat \rho_{\pi_k}|)$, since these $p$-values follow a single-step adjustment. 
For  $\ell=k+1,\ldots, M$, it is the application of the $\max$ operator (outside the square brackets) at each of those steps that guarantees the remaining step-down monotonicity. This approach can yield power improvements since the SD adjusted $p$-values are uniformly no larger than their SS counterparts. When $k=1$,  Definition (\ref{pSD}) simplifies to Westfall and Young's definition of SD adjusted $p$-values \citep[cf.][Eq. 15]{Ge-Dudoit-Speed:2003}.

Extending \citet[][Algorithm 4.1, pp. 116--117]{Westfall-Young:1993} and \citet[][Box 2]{Ge-Dudoit-Speed:2003}, Algorithm \ref{algorithm3} below shows how to compute the Monte Carlo version of the SD $p$-values defined in (\ref{pSD}). Here again the
number of resampling draws $B-1$ is assumed to be chosen so that $\alpha B$ is an integer, where $\alpha \in (0,1)$ is the user's desired $k$-FWER.

\begin{algorithm}[Step-down $k$-FWER-adjusted  Monte Carlo $p$-values]
\label{algorithm3}
\leavevmode 
\begin{enumerate}
{\normalfont

\item With the actual data, get the index values $\pi_1,\ldots,\pi_M$ that define the ordering $|\hat \rho_{\pi_1}| \ge |\hat \rho_{\pi_2}| \ge \ldots \ge |\hat \rho_{\pi_M}|$.

\item For $b=1,\ldots,B-1$, repeat the following steps:
  \begin{itemize}
 
 \item[(a)] generate an artificial data sample $\tilde{\mathbf Y}_b$  according to (\ref{ytil});

\item[(b)] compute the associated  matrix of correlations about the origin $\tilde {\boldsymbol \Gamma}_b =  [ \tilde \rho_{ij,b} ]_{N \times N}$ and let 
 $ \tilde{\boldsymbol \rho}_b = \text{vechs}(\tilde {\boldsymbol \Gamma}_b )$;

 \item[(c)] find the simulated successive maxima as 
\[
 \begin{split}
  \tilde \upsilon_{M,b}   & = | \tilde \rho_{\pi_M,b} |, \\
  \tilde \upsilon_{\ell,b} & = \max \big( \tilde \upsilon_{\ell+1,b},  | \tilde \rho_{\pi_\ell,b} | \big),  \text{ for } \ell=M-1,\ldots,k+1;
 \end{split}
\] 

 \item[(d)] refine these maxima as 
\[
 \begin{split}
  \tilde m^k_{1,b}      & =  \text{$k$-max}(|\tilde{ \boldsymbol \rho}_b|) ,   \\
  \tilde m^k_{\ell,b}   & =  \tilde m^k_{1,b} ,  \text{ for } \ell=2,\ldots,k,  \\  
  \tilde m^k_{\ell,b}   & = \min \big( \tilde m^k_{\ell-1,b}, \tilde \upsilon_{\ell,b} \big),  \text{ for } \ell=k+1,\ldots,M.
 \end{split}
\] 

\end{itemize}

\item Draw $u_b \sim \mathcal U(0,1),$ for $b=1,\ldots,B.$

\item For  $\ell=1,\ldots,M$,

\begin{itemize}

\item[(a)] create the pairs $(\tilde m^k_{\ell,1}, u_1 ),\ldots,(\tilde m^k_{\ell, B-1}, u_{B-1} ), (|\hat \rho_{\pi_\ell}|, u_B )$,  and compute   the lexicographic rank of $| \hat \rho_{\pi_\ell} | $ among the $\tilde m^k_{\ell,b}$'s as
 \[
 R^k_\text{SD}( | \hat \rho_{\pi_\ell} | ) = 1 + \sum_{b=1}^{B-1} \mathds{1} \left\{  | \hat \rho_{\pi_\ell} |  >  \tilde m^k_{\ell,b}   \right\}  + \sum_{b=1}^{B-1} 
 \mathds{1} \left\{  | \hat \rho_{\pi_\ell} |  =  \tilde m^k_{\ell,b}  \right\}   \mathds{1} \left\{ u_B > u_b  \right\};
\]

\item[(b)] compute the SD adjusted Monte Carlo $p$-value of $|\hat \rho_{\pi_\ell}| $ as
  \[
  \tilde p^k_\text{SD}(| \hat \rho_{\pi_\ell} |)  = \frac{B -  R^k_\text{SD}( | \hat \rho_{\pi_\ell} |) + 1}{B}.
  \]

\end{itemize}

\item Enforce monotonicity of the $p$-values by setting
\[
\begin{split}
  \tilde p^k_\text{SD}(| \hat \rho_{\pi_{\ell}} |) & \leftarrow   \tilde p^k_\text{SD}(| \hat \rho_{\pi_{\ell}} |), \quad \text{ for } \ell=1,\ldots,k, \\
  \tilde p^k_\text{SD}(| \hat \rho_{\pi_\ell} |) & \leftarrow   \max \left(    \tilde p^k_\text{SD}(| \hat \rho_{\pi_{\ell-1}} |), \tilde p^k_\text{SD}(| \hat \rho_{\pi_{\ell}} |) \right), \quad \text{ for } \ell=k+1,\ldots,M. 
\end{split}
\]

\item In terms of the original indices, the $p$-values $  \tilde p^k_\text{SD}(| \hat \rho_{\ell} |)$,  $\ell=1,\ldots,M$,
are recovered from $ \tilde p^k_\text{SD}(| \hat \rho_{\pi_1} |), \ldots, \tilde p^k_\text{SD}(| \hat \rho_{\pi_M} |) $ by reversing the mapping  $\ell \mapsto \pi_{\ell}$.

}
\end{enumerate}
\end{algorithm}

\begin{remark}
\normalfont
This algorithm requires more operations than its single-step counterpart and 
the bottleneck is obviously in Step 2. 
It is noteworthy, however, that for each replication the additional complexity in that step is linear in $M$. Indeed, relative to 
Algorithm \ref{algorithm2}, the number of additional comparisons needed in Steps 2-(c) and 2-(d) has order  $O(M)$, since each comparison only involves two elements.
\end{remark}

\begin{proposition}
\label{proposition5}
Under (\ref{model}) and Assumptions \ref{assumption1} and \ref{assumption2},  
Algorithm \ref{algorithm3} preserves strong control of the finite-sample $k$-FWER.
\end{proposition}

\begin{remark} 
\normalfont
Since  $ \tilde p^k_\text{SD}(| \hat \rho_{\ell} |)  \le \tilde p^k_\text{SS}(| \hat \rho_{\ell} |) $, 
given the same underlying randomization (i.e., the same artificial samples $\tilde{\mathbf Y}_b$, $b=1,\ldots,B-1$ and uniform draws $u_b$, $b=1,\ldots,B$) in Algorithms \ref{algorithm2} and \ref{algorithm3}, the SD adjustments will tend to be less conservative than their SS counterparts as the set $\mathcal M_0$ shrinks.
\end{remark}

\begin{remark}
\label{remark4.5} 
\normalfont
Observe also from the definition of $\text{$k$-max}(\cdot)$  that $\text{$k_1$-max}\bigl(|\tilde{ \boldsymbol \rho}|\bigr) \ge \text{$k_2$-max}\bigl(|\tilde{ \boldsymbol \rho}|\bigr)$  as long as $k_1 < k_2$. Comparing (\ref{pSS}) and (\ref{pSD}), it then becomes clear that $\tilde p^k_\text{SS}(| \hat \rho_{\ell} |) - \tilde p^k_\text{SD}(| \hat \rho_{\ell} |) \rightarrow 0$ as $k$ is increased, with the equality  $\tilde p^k_\text{SS}(| \hat \rho_{\ell} |) = \tilde p^k_\text{SD}(| \hat \rho_{\ell} |)$ holding when $k=M.$
\end{remark}

\subsection{Adjusted $p$-values for FDP control}

Following \citet[][Algorithm 4.1]{Romano-Wolf:2007}, control of the FDP in (\ref{FDP}) can be achieved by sequentially applying a $k$-FWER-controlling procedure, for $k=1,2,\ldots$, until a stopping rule indicates termination. Whether Algorithm \ref{algorithm2} or Algorithm \ref{algorithm3} is employed, it is important that the same underlying $\tilde{\mathbf Y}_b$, $b=1,\ldots,B-1$, and $u_b$, $b=1,\ldots,B$, be used for each $k$ to ensure coherence of the resulting $p$-values. This can be done simply by resetting the seed of the random number generator to the same value each time $k$ is changed, as illustrated in Algorithm \ref{algorithm4} below.

The user first specifies the FDP exceedance threshold  $\gamma \in [0,1)$ and chooses $B$ so that $\alpha B$ is an integer, where $\alpha \in (0,1)$ is the desired probability level.
In the following, $ \tilde p^{k}_{\diamond}(| \hat \rho_{\ell} |) $ represents either  $ \tilde p^k_\text{SS}(| \hat \rho_{\ell} |)$ or $ \tilde p^k_\text{SD}(| \hat \rho_{\ell} |) $ computed according to Algorithms \ref{algorithm2} and \ref{algorithm3}, respectively. Let {\tt seed} be a value that will be used to set the seed of the random number generator.\footnote{In the R programming language this can be done with the command  {\tt set.seed({\tt seed})} where {\tt seed}=8032, for example.}

\begin{algorithm}[FDP-adjusted  Monte Carlo $p$-values]
\label{algorithm4}
\leavevmode 
\begin{enumerate}
{\normalfont

\item  Initialize: $k \leftarrow 1$. Set the seed of the random number generator to {\tt seed}.

\item Apply the $k$-FWER-controlling procedure to get the $p$-values $\tilde p^{k}_{\diamond}(| \hat \rho_{1} |), \ldots, \tilde p^{k}_{\diamond}(| \hat \rho_{M} |)  $, count the number of hypotheses rejected $R_k = \sum_{\ell=1}^M \mathds{1} \big \{ \tilde p^{k}_{\diamond}(| \hat \rho_{\ell} |) \le \alpha   \big\}$,  and store these values. 

\item If $k > \gamma(R_k + 1)$, stop. If this occurs because $\gamma=0$, then the FDP-adjusted $p$-values are $\tilde p^{\gamma}_{\diamond}(| \hat \rho_{\ell} |) = \tilde p^{k^{\ast}}_{\diamond}(| \hat \rho_{\ell} |)$, for $\ell=1,\ldots,M,$ where $k^{\ast}=1.$\footnote{Recall that controlling the FDP with $\gamma=0$ is  equivalent to controlling the $1$-FWER.} 
If stopping occurs with $\gamma > 0$, conclude that FDP-adjusted $p$-values cannot be produced.\footnote{Algorithm 4.1 in \citet{Romano-Wolf:2007} does not consider the possibility that $1 > \gamma(R_1+1)$.}

\item While $k \le \gamma(R_k + 1)$ repeat the following steps:
\begin{itemize}
 
    \item[(a)]  increment: $k \leftarrow k+1$; reset the seed of the random number generator to {\tt seed};

    \item[(b)] apply the $k$-FWER-controlling procedure to get $\tilde p^{k}_{\diamond}(| \hat \rho_{1} |), \ldots, \tilde p^{k}_{\diamond}(| \hat \rho_{M} |)  $, compute 
    $R_k = \sum_{\ell=1}^M \mathds{1} \big \{ \tilde p^{k}_{\diamond}(| \hat \rho_{\ell} |) \le \alpha   \big\}$, and store these values.

\end{itemize}

\item Upon termination of the  while loop (i.e., as soon as $k > \gamma(R_k + 1)$), the FDP-adjusted $p$-values are $\tilde p^{\gamma}_{\diamond}(| \hat \rho_{\ell} |) = \tilde p^{k^{\ast}}_{\diamond}(| \hat \rho_{\ell} |)$, for $\ell=1,\ldots,M,$ where $k^{\ast}=k-1.$

}
\end{enumerate}
\end{algorithm}

\begin{proposition}
\label{proposition6}
Under (\ref{model}) and Assumptions \ref{assumption1} and \ref{assumption2}, the $p$-values produced by 
Algorithm \ref{algorithm4} have control of the finite-sample FDP in the sense that 
$\Pr \left( F_{k^{\ast}} / R_{k^{\ast}}  > \gamma \, \vert \, \bigcap_{\ell \in \mathcal M_0}  H_{\ell}  \right) \le \alpha$, where 
$R_{k^{\ast}} = \sum_{\ell=1}^M \mathds{1} \big \{ \tilde p^{{k^{\ast}}}_{\diamond}(| \hat \rho_{\ell} |) \le \alpha   \big\}$ is the 
total number of rejections and 
$F_{k^{\ast}}$ is the number of false rejections made by the underlying $k^{\ast}$-FWER-controlling procedure.
\end{proposition}

\begin{remark} 
\normalfont
Obviously the FDP-adjusted $p$-values $\tilde p^{\gamma}_{\diamond}(| \hat \rho_{\ell} |)$  produced by 
Algorithm \ref{algorithm4} also become more conservative as the cardinality of $\mathcal M_0$ decreases, just like  $ \tilde p^k_\text{SS}(| \hat \rho_{\ell} |)$ and $ \tilde p^k_\text{SD}(| \hat \rho_{\ell} |) $ upon which they rest.
\end{remark}

For large $M$, Algorithm \ref{algorithm4} can be exceedingly slow, since it finds $k^*$ by progressing sequentially starting with $k=1$, then $2$, and so on.
The monotonicity of the underlying $k$-FWER-controlling procedure can be further exploited to 
achieve important speed gains via a  bisection method, as follows.

\begin{algorithm}[FDP-adjusted  Monte Carlo $p$-values via bisection]
\label{algorithm5}
\leavevmode 
\begin{enumerate}
{\normalfont

\item  Initialize: $k_l \leftarrow 1$ and $k_r \leftarrow M$. 

\item While $k_r - k_l > 1$, repeat the following steps:

\begin{itemize}
 
 \item[(a)] set $k_m = \lfloor (k_l + k_r)/2 \rfloor$, where $\lfloor x \rfloor$ gives the greatest integer less than or equal to $x$,  and reset the seed of the random number generator to {\tt seed};
 
 \item[(b)] apply the $k_m$-FWER-controlling procedure to get $\tilde p^{k_m}_{\diamond}(| \hat \rho_{1} |), \ldots, \tilde p^{k_m}_{\diamond}(| \hat \rho_{M} |)  $ and compute 
    $R_{k_m} = \sum_{\ell=1}^M \mathds{1} \big \{ \tilde p^{k_m}_{\diamond}(| \hat \rho_{\ell} |) \le \alpha   \big\}$;

 \item[(c)] if $k_m \le \gamma(R_{k_m} + 1)$ then set $k_l = k_m$, otherwise set $k_r = k_m$.
 
 \end{itemize}

\item Proceed with Algorithm \ref{algorithm4}, replacing $k \leftarrow 1$ in Step 1 by $k \leftarrow k_l$.

}
\end{enumerate}
\end{algorithm}
The bisection method, known for its numerical robustness, is particularly well-suited to handle the discontinuities within the underlying step function $R_k$.
Steps 1 and 2 isolate the largest  value $k_l$ such that $1 \le k_l \le k^* \le M$ and $\gamma(R_{k_l} + 1) = \gamma(R_{k^{\ast}} + 1)$.
This is achieved by successively narrowing down the interval in which $k_l$ lies through a process of halving at each iteration of Step 2.
The call in Step 3 to Algorithm \ref{algorithm4}, initialized with $k_l$, then returns 
$\tilde p^{\gamma}_{\diamond}(| \hat \rho_{\ell} |)$, $\ell=1,\ldots,M,$ with at most two iterations of the while loop in that algorithm.

\subsection{Covariance matrix estimators}

The next step in the construction of the proposed multiple-testing regularized estimators is to set to zero the statistically insignificant entries of the sample correlation matrix $\hat {\boldsymbol \Gamma}  = [ \hat \rho_{ij} ]_{N \times N}$,  defined previously.
Let $ \tilde p^{\diamond}_{\diamond}(| \hat \rho_{ij} |) $ represent either  $ \tilde p^k_\text{SS}(| \hat \rho_{ij} |)$, $ \tilde p^k_\text{SD}(| \hat \rho_{ij} |) $, $ \tilde p^\gamma_\text{SS}(| \hat \rho_{ij} |)$, or $ \tilde p^\gamma_\text{SD}(| \hat \rho_{ij} |) $.
A correlation $p$-value matrix corresponding directly to $\hat {\boldsymbol \Gamma} $ is given by $\text{fill}\big(\tilde p^{\diamond}_{\diamond}(| \hat \rho_{1} |), \ldots, \tilde p^{\diamond}_{\diamond}(| \hat \rho_{M} |)  \big)$ with the diagonal elements set to zero.\footnote{These zero diagonal values are irrelevant since the diagonal elements of ${\boldsymbol \Sigma}$ are not tested.}

The associated correlation matrix estimator $\hat {\boldsymbol \Gamma}^{\diamond}_{\diamond}  = [ \hat \rho^{\diamond}_{\diamond, ij} ]_{N \times N}$ has entries given by
\begin{equation}
\hat \rho^{\diamond}_{\diamond,ij}  = \hat \rho_{ij} \mathds{1} \big \{ \tilde p^{\diamond}_{\diamond}(| \hat \rho_{ij} |) \le \alpha   \big\}. \quad  \label{adjust}
\end{equation}
These adjustments to the sample correlation matrix are made  for $j=1,\ldots,N-1$ and $i=j+1,\ldots,N$; the diagonal elements of $\hat {\boldsymbol \Gamma}^{\diamond}_{\diamond} $ are obviously set as $\hat \rho^{\diamond}_{\diamond,ii} =1$; and symmetry is imposed by setting $\hat \rho^{\diamond}_{\diamond,ij} = \hat \rho^{\diamond}_{\diamond,ji}.$
In contrast to the universal threshold critical value  used in (\ref{bps}) for each of the pairwise correlations, note that the  $p$-values used in (\ref{adjust}) are fully data-driven and adapt to the variability of individual entries of the correlation matrix.

Proceeding to the shrinkage step in (\ref{shrink}) of the BPS approach with  $\hat {\boldsymbol \Gamma}^{\diamond}_{\diamond} $ instead of $\hat {\boldsymbol \Gamma}_{\text{BPS}} $ yields the positive definite correlation matrix estimator
$\hat {\boldsymbol \Gamma}^{\diamond}_{\diamond}(\xi^*) = \xi^* \mathbf I_N + (1-\xi^*) \hat {\boldsymbol \Gamma}^{\diamond}_{\diamond},$
where $\xi^* = \arg \min_{\xi_0 \le \xi \le 1} \left\lVert  \hat{\boldsymbol \Gamma}^{-1}_0   - \hat {\boldsymbol \Gamma}_{\diamond}^{\diamond^{ -1}}(\xi) \right\rVert^2_F,   $
and the associated covariance matrix estimator
$\hat {\boldsymbol \Sigma}^{\diamond}_{\diamond}  (\xi^*) =      \hat {\mathbf D}^{1/2} \hat {\boldsymbol \Gamma}^{\diamond}_{\diamond}(\xi^*)  \hat {\mathbf D}^{1/2}.$

\subsection{Unknown location parameters}

Observe that the null distribution generated according to (\ref{ytil}) depends on ${\boldsymbol \mu}^{\ast}$ in (\ref{model}) only through the subtractive transformations $\mathbf y_t = \mathbf r_t  - \boldsymbol \mu^{\ast}$, $t=1,\ldots,T$. When the values comprising ${\boldsymbol \mu}^{\ast}$ are unknown, it will be assumed that  they can be estimated consistently.
Let $ \hat {\boldsymbol \mu}_T $  denote the equation-by-equation estimator of ${\boldsymbol \mu}^{\ast}$ obtained from a sample of size $T$.
For instance, this could be the sample mean. With $ \hat {\boldsymbol \mu}_T $ in hand, the Monte Carlo procedures can proceed as before except that $\hat{\mathbf y}_t = \mathbf r_t - \hat{\boldsymbol \mu}_T $ replaces $\mathbf y_t = \mathbf r_t - {\boldsymbol \mu}^{\ast}$.

Let $\hat \rho_{ij}( \hat {\boldsymbol \mu}_T )$, $1 \le j < i \le N$, denote the correlations estimated on the basis of $\hat{\mathbf Y}  = \left[\hat{\mathbf y}_1,\ldots,\hat{\mathbf y}_T  \right]^{\prime}$. Since $ \hat {\boldsymbol \mu}_T \stackrel{p}{\rightarrow} {\boldsymbol \mu}^{\ast}$, Slutsky's theorem ensures that  $\hat{ \mathbf  Y}  \stackrel{d}{\rightarrow}    \mathbf  Y  \stackrel{d}{=} \tilde{\mathbf Y}$, under $H_0$, as $T \rightarrow \infty$. From  \citet[][Section 4]{Zhu-Neuhaus:2000} and \citet[][Theorem 3]{Toulis-Bean:2021}, it then follows  that 
the Monte Carlo $p$-values computed according to Algorithm \ref{algorithm1}  are asymptotically valid in the sense that $\Pr(   \tilde p_\text{U}(| \hat \rho_{ij}( \hat {\boldsymbol \mu}_T ) |)  \le \alpha   \, \vert \, H_0  ) = \alpha + o_p(1)$,  for  $1 \le j < i \le N$, as $T \rightarrow \infty$. An immediate consequence is that the adjusted $p$-values computed according to Algorithms \ref{algorithm2}--\ref{algorithm4} are also  asymptotically valid, since they all rest on Algorithm \ref{algorithm1}.

\section{Simulation experiments}

This section examines the performance of the proposed Monte Carlo regularized covariance estimators, with the BPS approach serving as the natural benchmark for comparisons. 
The simulation experiments are designed to resemble the empirical application presented in the next section.
For this purpose the data-generating process for daily returns $\mathbf r_t=(r_{1,t},\ldots,r_{N,t})^{\prime}$ is specified as a CCC model \citep{Bollerslev:1990} of the form
\vspace{-0.5cm}
\begin{align*}
\mathbf r_t & = \boldsymbol \mu + \boldsymbol \Sigma_t^{1/2} \mathbf z_t, \\
\boldsymbol \Sigma_t & = \mathbf D_t^{1/2} \boldsymbol \Gamma \mathbf D_t^{1/2},
\end{align*}
where $\mathbf  D_t = \text{diag}(\sigma_{1,t}^2,\ldots, \sigma_{N,t}^2)$ is an $N \times N$ diagonal matrix comprising the time-$t$ conditional variances, and $ \boldsymbol \Gamma$ is a constant conditional correlation (CCC) matrix. 
The vector $ \boldsymbol \mu $ is set to zero, but this is not assumed known and the multiple testing procedures are applied with  $\mathbf y_t = \mathbf r_t  - \bar{\mathbf r}$,  where $\bar{\mathbf r}$ is the vector of sample means (cf. Section 4.6).
The conditional variances appearing in $\mathbf D_t$ evolve according to the standard $\text{GARCH}(1,1)$ model $
\sigma^2_{i,t} = \theta_0 + \theta_1 r_{i,t-1}^2 + \theta_2 \sigma^2_{i, t-1} $ with common parameters across assets set as $\theta_0 = 0.01$, $\theta_1=0.1$, and $\theta_2=0.85$. 
These are typical values found with daily returns data.
The innovation terms $ z_{i,t}$ are i.i.d.  according either to the standard normal distribution, or the standardized $t$-distribution with 12 or 6 degrees of freedom.

The correlation structure is defined as follows. Given a value $\delta$, $0 \le \delta \le 1$, the vector $\mathbf c = (c_1,\ldots,c_N)^{\prime}$ is filled in with $N_c= \lfloor \delta N \rfloor$ non-zero elements drawn from the triangular distribution on $[0,1]$ with mode at $1$, and the remaining $N-N_c$ elements are set to zero. The positions of the zero and non-zero elements within  $\mathbf  c $ are random. Following BPS, this vector is then used to obtain a well-defined correlation matrix as $\boldsymbol \Gamma = \mathbf I_N + \mathbf c \mathbf c^{\prime} - \text{diag}( \mathbf c \mathbf c^{\prime} ).$ The complete null hypothesis in (\ref{null}) is thus represented by $\delta=0$ and increasing the value of $\delta$ towards 1 results in a smaller set $\mathcal M_0$ of true hypotheses. With $\boldsymbol \Gamma$ in hand, the unconditional covariance matrix is then found as ${\boldsymbol \Sigma} =  \mathbf D^{1/2} \boldsymbol \Gamma \mathbf D^{1/2},$
where $\mathbf  D = \text{diag}(\sigma_{1}^2,\ldots, \sigma_{N}^{2})$ comprises the GARCH-implied unconditional variances given by
 $\sigma_i^2=\theta_0/(1-\theta_1-\theta_2)$, for $i=1,\ldots,N.$ Vector sizes $N=25$, $100$, and $500$, and sample sizes $T=63$, $126$, and $252$ are considered.

Following BPS, two choices are used to complete the definition in (\ref{crit}) of the universal critical value: (i) $f(N) = N^2$, which results in a more conservative adjustment; and  (ii)  $f(N) = N(N-1)/2$, representing the well-known Bonferroni rule.
The results based on these choices are labelled $\text{BPS}_a$    and $\text{BPS}_b$, respectively.

In the application of the $k$-FWER procedures, reported as $\text{SS}_{k}$ and $\text{SD}_{k}$, two heuristic rules are considered:  (i) $k=\lfloor \log{M} \rfloor$ and (ii) $k=\lfloor \sqrt{M} \rfloor$. These choices scale the allowable number of {Type I} errors with the number of hypotheses being tested, allowing for a more permissive approach to error control. 
The exceedance threshold in the FDP procedures, whose results are reported as $\text{SS}_{\gamma}$ and $\text{SD}_{\gamma}$, is set at $\gamma=0.1.$ The number of replications for the Monte Carlo procedures is established with $B=100$ and all the tests are conducted with  $\alpha = 0.05.$

\subsection{Numerical results}

Table 1 reports the empirical rejection rates of the FWER procedures.
Panels A and B report the empirical FWER and Panel C reports the empirical average power, defined as the average expected number of correct rejections among all false null hypotheses 
\citep[][Section 2.1]{Bretz-Hothorn-Westfall:2010}. 
Under the complete null hypothesis ($\delta=0$) the  SS and SD test procedures have identical rejection rates, indicated on the lines labelled SS/SD in Panel A.
It is seen that the  SS (SD) procedure does a good job at keeping the FWER close to the desired $5\%$ value when $\delta=0$.
This case provides a comparison benchmark for the FWER results in Panel B when $\delta$ is increased to $0.9$. 
These tests are seen in Panel B to also maintain control of the FWER in the strong sense. 
Comparing Panels A and B shows that the SS and SD test procedures become more conservative when the number of true null hypotheses diminishes ($\delta$ increasing), as expected from the developed theory.

The BPS method also achieves good control of the FWER under normality. 
And just like the SS and SD procedures, the BPS tests are seen  in Panel B to also become conservative as $\delta$ reaches 0.9.
However when the  error terms are non-normal, the BPS approach tends to spuriously over-reject. 
Table 1 reveals that the BPS over-rejection problem worsens as: (i) the degree of tail heaviness increases (from $t_{12}$ to $t_6$ errors), (ii) $N$ increases, and (iii) $T$ increases. 
The most egregious instance occurs with $T=252$ where the FWER with $\text{BPS}_a$ and $\text{BPS}_b$ attains nearly $100\%$ when $N=500$ under $t_6$ errors.
This makes clear that finite-sample FWER control with the BPS thresholding estimator based on $c_{\alpha}(N)$ in (\ref{crit}) is heavily dependent on normality, even as the sample size $T$ increases. 
Therefore in order to ensure a fair comparison, all the power results for the BPS approach are  based on strong FWER-adjusted critical values.\footnote{  
Of course, the FWER-adjusted BPS method is not feasible in practice. It is merely used here as a benchmark for the SS and SD procedures.}

Panel C of Table 1 shows the  power of the three FWER procedures when $\delta=0.9$.
Note that  $\text{BPS}_a$ and $\text{BPS}_b$ have identical size-adjusted power, reported on the lines labelled $\text{BPS}_a/\text{BPS}_b$.
The results in Table 1 show that SS tends to be less powerful than SD, whose average power is quite close to that of BPS, especially as the sample size grows and tail heaviness increases. 
As expected, power is seen to increase with $T$, and to decrease as  $N$ grows large and  tail heaviness increases.

Table 2 reports the empirical rejection rates of the $k$-FWER procedures when $\delta=0.9.$ 
In Panel A, the empirical probability of making $k$ or more false rejections is effectively zero, corroborating Propositions \ref{proposition4} and \ref{proposition5}.
Table 2 reveals that the average power of the $k$-FWER procedures is higher when compared to the SS and SD results presented in Table 1.
In fact, in Panel B, both $\text{SS}_{k}$ and $\text{SD}_{k}$ exhibit increasing power, with the gap between them narrowing as $k$ increases (cf. Remark \ref{remark4.5}).

The empirical exceedance probabilities and average power of the FDP procedures with $\gamma=0.1$ are reported in Table 3, again for the case where $\delta=0.9.$ 
Panel A shows that the procedures effectively control the FDP criterion by ensuring that the empirical probability of having a proportion of false discoveries exceed $\gamma$ is zero. From Panel B, it is clear that $\text{SS}_{\gamma}$ and $\text{SD}_{\gamma}$ tend to have the same average power as $M$ grows large.
Compared to the $k$-FWER results in Table 2, the FDP procedures appear to reject more false hypotheses, particularly when there is a large number $M$ of hypotheses being tested.

\section{Application to portfolio optimization}

The proposed multiple testing procedures are further illustrated in this section with an application to Markowitz portfolio optimization using  
daily stock return data downloaded from the Center for Research in Security Prices (CRSP) for the period starting on 
January 1, 2004 and ending on December 31, 2021 (4531 observations).\footnote{Only U.S.-based common stocks as identified via CRSP share codes 10 and 11 from the NYSE, NYSE MKT, and NASDAQ exchanges are considered. \citet[][Ch. 3]{Scheuch-Voigt-Weiss:2023} show 
how to access, filter, and download daily CRSP data in small batches  through a  Wharton Research Data Services (WRDS) connection  using {R}.} 
The considered problem is that of an investor whose objective is to hold a global minimum variance (GMV) portfolio based on $N \in \{25, 100, 500  \}$ stocks over the next $21$ days, which is considered a ``month." 
As \citet{Engle-Ledoit-Wolf:2019} and \citet{DeNard-Engle-Ledoit-Wolf:2022} argue in a similar context, the accuracy of the covariance matrix estimators should be evaluated primarily by the out-of-sample standard deviation achieved by the GMV portfolios. 
Nonetheless, there is also a significant interest in exploring additional performance measures.
This interest persists even when considering that more realistic portfolio management scenarios would involve limitations on maximum position, industry position, factor exposure, etc. \citep{DeNard-Engle-Ledoit-Wolf:2022}.

Let $t_{\text{b}}$ refer to the day when the portfolio is initially formed and subsequent rebalancing days. 
On those days the investor uses the returns from the past $L \in \{ 63, 126, 252 \}$ days to obtain $\hat{ \boldsymbol \Sigma }_{t_{\text{b}}}$, an estimate of the covariance matrix. 
The investor then uses $\hat{ \boldsymbol \Sigma }_{t_{\text{b}}}$ to find the GMV portfolio weights 
$ \hat {\boldsymbol \omega}_{t_{\text{b}} } = (\omega_{1,t_{\text{b}}},\ldots,\omega_{N,t_{\text{b}}})^{\prime}$ 
by solving the problem
\begin{equation}
\begin{split}
\min_{\boldsymbol \omega }  &\;\; \boldsymbol \omega^{\prime}{ \boldsymbol \Sigma }  \boldsymbol \omega \\
\text{subject to}  		    &\;\; \boldsymbol \iota^{\prime} \boldsymbol \omega  = 1, \\ 
							&\;\;\;\; \boldsymbol \omega  \ge  \boldsymbol 0,  \label{GMV}
\end{split}
\end{equation}
with the assignment $\boldsymbol \Sigma \leftarrow \hat{ \boldsymbol \Sigma }_{t_{\text{b}}}$, and where $\boldsymbol \iota$ denotes an $N \times 1$ vector of ones. 
The second constraint in (\ref{GMV}) restricts each $\omega_{i,t_{\text{b}}}$ to be non-negative, meaning that short sales are prohibited. 
When short sales are allowed, the analytical solution to the GMV problem is
$ \hat{ \boldsymbol \Sigma }_{t_{\text{b}} } ^{-1} \boldsymbol \iota / ( \boldsymbol \iota^{\prime}  \hat{ \boldsymbol \Sigma }_{t_{\text{b}} } ^{-1}  \boldsymbol \iota) $.
With the short-selling restriction, the GMV problem is a quadratic optimization problem that can be solved numerically.\footnote{See \citet[][Section 17.5]{Scheuch-Voigt-Weiss:2023} for a more detailed description of constrained portfolio optimization along with R code.} Restricting short sales is equivalent to shrinking towards zero the larger elements of the covariance matrix that would otherwise imply negative weights  \citep{Jagannathan-Ma:2003}, offering an interesting point of comparison.

With each choice of estimation window length $L$, the initialization $t_{\text{b}}  \leftarrow 252$  corresponds to  January 3, 2005, when the portfolio is first formed.
This portfolio is then held for $21$ days and the resulting realized out-of-sample portfolio returns are $\hat {\boldsymbol \omega}_{t_{\text{b}} }^{\prime}   \mathbf r_{\tau}  $, for $\tau = t_{\text{b}}+1,\ldots,t_{\text{b}} + 21.$
After its initial formation, the portfolio is rebalanced on days $t_{\text{b}} \leftarrow t_{\text{b}} + 21$. 
This process yields 203 $t_{\text{b}}$'s  and 4263 out-of-sample returns.
Rebalancing consists of finding new  GMV weights by solving (\ref{GMV})  with the updated covariance matrix estimate based on the returns from the last $L$ days.
Following \citet{Engle-Ledoit-Wolf:2019}, the investment universe on days $t_{\text{b}}$ is obtained by finding the $N$ largest capitalization stocks with a complete return history over the most recent $L$ days and a complete return future over the next 21 days. This way the composition of the investment universe evolves with each rebalancing day.

To obtain $\hat{ \boldsymbol \Sigma }_{t_{\text{b}}}$, it is natural to first consider the sample covariance matrix. Of course, the other choices considered include
the  approaches based on FWER control ($\text{BPS}_a$, $\text{BPS}_b$, $\text{SS}$, $\text{SD}$),  the  approaches based on $k$-FWER control ($\text{SD}_{\lfloor \log{M} \rfloor}$ and
 $\text{SD}_{\lfloor \sqrt{M} \rfloor}$), and the approaches based on FDP control ($\text{SS}_{\gamma}$ and $\text{SD}_{\gamma}$, with $\gamma=0.1$), which proceed by testing the significance of the $M \in \{300, 4950, 124750 \}$ distinct covariances in  the rolling-window scheme.\footnote{The nominal levels are set to $\alpha = 5\%$ and $B-1=99$ resampling draws are used in the computation of the Monte Carlo $p$-values.} 
The effectiveness of those approaches is further assessed by comparing them to portfolios based on two shrinkage covariance matrix estimators:
(i) the linear shrinkage (LS) estimator of \citet{Ledoit-Wolf:2004}, which shrinks the sample covariance matrix towards the identity matrix; and (ii) the 
non-linear shrinkage (NLS) estimator proposed by  \citet{Ledoit-Wolf:2015}.\footnote{Specifically, the Ledoit-Wolf shrinkage covariance matrix estimates are computed with the {\tt linshrink\_cov} and {\tt nlshrink\_cov} commands available with the R  package `nlshrink' \citep{Ramprasad:2016}.} 

The performance evaluation also includes the  equally weighted (EW) portfolio, which bypasses (\ref{GMV}) altogether and simply sets $\hat \omega_{i,t_{\text{b}}}=1/N,$ $i=1,\ldots, N.$ This naive portfolio is a standard benchmark for comparisons; see \citet{DeMiguel-Garlappi-Uppal:1999} and \citet{Kirby-Ostdiek:2012}, among others. The final strategy considered is the volatility timing (VT) portfolio with weights $\hat \omega_{i,t_{\text{b}}} = \hat\sigma^{-2}_{i,t_{\text{b}}}  / \sum_{i=1}^N \hat\sigma^{-2}_{i,t_{\text{b}}},$ $i=1,\ldots, N,$ which was suggested by \citet{Kirby-Ostdiek:2012} as a competitor to  EW. The VT portfolio can be seen as an aggressive form of shrinkage that sets to zero all the off-diagonal elements of the sample covariance matrix.

Note that there is no estimation risk associated with the EW strategy, which helps reduce the portfolio turnover.
On the contrary, active strategies that generate high turnover will suffer more in the presence of transaction costs. 
To see this, note that for every dollar invested in the portfolio at time $t_{\text{b}}$, there are 
$ \hat \omega_{i,t_{\text{b}}} \prod_{\tau=t_{\text{b}}+1}^{t} (1+ r_{i,\tau}) $  dollars  invested in asset $i$ at time $t$, for $t > t_{\text{b}}$ and as long as the portfolio is not rebalanced.
Hence, at any time $t$ until the the next rebalancing occurs, the actual weight of asset $i$ in the portfolio is
\[
\omega^{\ast}_{i,t} =  \frac{ \hat \omega_{i,t_{\text{b}}}\prod_{\tau=t_{\text{b}}+1}^{t} (1+ r_{i,\tau}) } {  \sum_{i=1}^N \hat \omega_{i,t_{\text{b}}} \prod_{\tau=t_{\text{b}}+1}^{t} (1+ r_{i,\tau}) }. 
\]
When rebalancing occurs, the portfolio turnover can be defined as
$\text{TO}_{t} = \sum_{i=1}^N \vert  \hat \omega_{i, t } -  \omega^{\ast}_{i,t}   \vert, $
where $\hat \omega_{i, t }$ is the updated weight for asset $i$ at rebalancing time $t= t_{\text{b}}$.
Denoting by $\kappa$ the transaction cost in proportion to the amount of wealth invested, the proportional cost of rebalancing all the portfolio positions is $\text{TC}_{t} = \kappa \text{TO}_{t}$ when $t= t_{\text{b}}$.  
Therefore, starting with $t_{\text{b}}  \leftarrow 252$  and a normalized initial wealth of $W_{t_{\text{b}}} = \$1$ on that first day, wealth subsequently evolves  according to 
\[
W_{t+1} = \left\{   \begin{array}{ll}

                        W_{t} \left( 1 +  \sum_{i=1}^N  \hat \omega_{i, t } r_{i,t+1} \right) \left( 1- \text{TC}_{t} \right), & \text{ when } t = t_{\text{b}}, \\[2.0ex]                                   

                        W_{t} \left( 1 +  \sum_{i=1}^N   \omega^{\ast}_{i, t } r_{i,t+1} \right) , & \text{ when } t \neq t_{\text{b}},

\end{array} \right.
\]
for $t = 252,\ldots,T-1$, with the updating rule $t_{\text{b}} \leftarrow t_{\text{b}} + 21$ to determine the rebalancing days. 
The out-of-sample portfolio return net of transaction costs is then given by $(W_{t+1}- W_{t})/W_{t}$.
Following \citet{DeNard-Engle-Ledoit-Wolf:2022}, $\kappa$ is set to 5 basis points to account for  transaction costs.\footnote{\citet{French:2008} estimates the  cost of trading stocks listed on the NYSE, AMEX, and NASDAQ, including total commissions, bid-ask spreads, and other costs investors pay for trading services. He finds that these costs have dropped significantly over time ``from 146 basis
points in 1980 to a tiny 11 basis points in 2006.''}

For each portfolio strategy, the following out-of-sample performance metrics are computed: (i) AV, the average return (annualized by multiplying by 252) in percent; (ii) SD, the standard deviation of returns (annualized by multiplying by $\sqrt{252}$) in percent;  (iii) IR, the information ratio  given as the annualized mean divided by the annualized standard deviation; (iv)  TO, the average turnover; (v) MDD, the maximum drawdown in percent over the trading period;\footnote{A drawdown refers to the loss in the investment's value from a peak to a trough, before a new peak is attained. MDD is thus an indicator of downside risk.} 
and (vi) TW, the terminal wealth in dollars at the end of the trading period.
Tables 4 and 5 present the results for $N=100$ and $500$, respectively. (The results with $N=25$ are in the Supplementary material.)
When $N > L$, the sample covariance matrix is singular. Those cases are indicated with n/a in the tables.

Since a major objective of this paper is to show that the BPS approach can be improved upon, the statistical significance of the  standard deviation differential between a given portfolio strategy and $\text{BPS}_b$ is assessed using the two-sided $p$-value of the prewhitened $\text{HAC}_{\text{PW}}$ method described in \citet{Ledoit-Wolf:2011} for testing the null hypothesis of equal standard deviations.\footnote{Note that these $p$-values assess the significance of pairwise differences; they no dot account for the multiplicity of tests.} This is done for each $(N, L)$ combination and whether short sales are allowed or not.
Observe that the performance of the EW portfolio is influenced by the choice of $L$, since  the composition of the investment universe on rebalancing days depends on the last $L$ \emph{and}  next 21 days.
The main findings with respect to the out-of-sample standard deviation  can be summarized as follows:

\begin{enumerate}

\item[$\bullet$] Compared to $\text{BPS}_b$, the performance of the EW portfolio is always statistically worse. In fact, all the GMV portfolios deliver lower standard deviations than the EW portfolio. An exception occurs when short selling is allowed with $N=100$ and $L=126$ (Table 4, Panel A), where the sample covariance matrix misbehaves and performs worse than EW. 

\item[$\bullet$] In Table 4, the performance of ``Sample'' generally improves as $L$ increases, achieving statistically lower standard deviations than the $\text{BPS}_b$ strategy. As the investment universe expands to $N=500$ in Table 5, the sample covariance matrix is nowhere available.

\item[$\bullet$] The VT portfolio consistently performs worse than $\text{BPS}_b$, which suggests that this form of shrinkage  is too extreme.

\item[$\bullet$] Turning to the multiple testing strategies, SS and SD are also seen to be too strict, never outperforming $\text{BPS}_b$ by a statistically significant margin. 
Observe also  that $\text{BPS}_a$ does not generally  perform any better than $\text{BPS}_b$.

\item[$\bullet$] The $k$-FWER procedure based on the $\lfloor \sqrt{M} \rfloor$ rule is either comparable or better than $\text{BPS}_b$. In particular when short selling is prohibited, 
$\text{SS}_{\lfloor \sqrt{M} \rfloor}$ and $\text{SD}_{\lfloor \sqrt{M} \rfloor}$ outperform  $\text{BPS}_b$ by a statistically significant margin.

\item[$\bullet$] As $N$ increases, the two FDP procedures  tend to outperform the other multiple testing strategies, and even more so under the short selling restriction. 
Indeed, Panels B of Tables 4 and 5 reveal that
the performance of  $\text{SS}_{\gamma}$ and $\text{SD}_{\gamma}$ over $\text{BPS}_b$ is always statistically  better at the 1\% level.

\item[$\bullet$] When using the procedures for either $k$-FWER or FDP control, it is seen, as expected, that proceeding with SS or SD tends to yield similar results as $N$ increases. 

\item[$\bullet$] Consistently, the lowest standard deviation is achieved either by LS or  NLS. 

\end{enumerate}

While minimizing volatility is a primary goal for a GMV investor, examining the other metrics in Tables 4 and 5 offers a more comprehensive view of risk, return, and overall performance. 
For instance, the TW columns reveal that wealth accumulated with LS and NLS strategies tends to be lower, especially when short selling is allowed. This can also be appraised from the AV columns in Tables 4 and 5, which show the tendency of LS and NLS to  yield lower mean returns in comparison to the other portfolio strategies.

The EW portfolio generally results in the greatest accumulated wealth, owing to its near zero turnover. As expected, however, this payoff involves greater risk as can be seen from the SD and MDD columns. Indeed, the standard deviation under the EW strategy is seen everywhere in Tables 4 and 5 to be statistically larger at the 1\% level than under $\text{BPS}_b$. 
And the MDD, which indicates the most significant loss sustained during the trading period, is exacerbated under the EW strategy.

The multiple testing strategies appear as ``Goldilocks" solutions, striking a balance between  risk and reward. In particular, when $N=100$  
and $N=500$ and short selling is prohibited (Panel B in Tables 4 and 5), the stricter SS strategy tends to result in greater end-of-period  wealth when compared to the other multiple testing strategies, as seen in the TW columns. 
This is further corroborated by comparing SS with VT in Panel B of Table 5, where terminal wealth with the even more stringent VT portfolio is higher than with the SS portfolio.
On the other hand, the MDD columns reveal that the  more lenient FDP procedures ($\text{SS}_{0.1}$ and $\text{SD}_{0.1}$) generally offer better protection against downside risk.

Further insight into these results can be gleaned from Figures 1--3, which show the proportion of  correlations declared statistically significant by the multiple testing procedures each time the portfolio is rebalanced, given an estimation window of length $L=252$ days and $N=500$ assets. 
Figure 1 shows the results for the FWER procedures, Figure 2 for the $k$-FWER procedures, and Figure 3 for the FDP procedures.
The solid line in Figure 1 shows that among the FWER procedures,   $\text{BPS}_b$ declares the greatest number of non-zero correlations, followed by SD (dotted line), and then SS (dashed line). A comparison with Figure 2 reveals that the number of significant correlations found with $\text{BPS}_b$ generally falls in between $\text{SS}_{11}$ and $\text{SS}_{353},$ except for the period 
from March 2020 to March 2021 when $\text{BPS}_b$ is above $\text{SS}_{353}$. 
And from Figure 3 it is clear that $\text{SS}_{0.1}$ is most lenient in declaring non-zero correlations.

Finally note that these figures provide a gauge of the portfolio's overall diversification profile under a given multiple testing criterion.
An improved profile is indicated when the lines dip; i.e., in periods when there are relatively fewer significant correlations among these large-cap stocks whose prices typically move in tandem.
It is interesting to observe that $\text{BPS}_b$ surges above $\text{SS}_{353}$ in March 2020, precisely when the World Health Organization declared the COVID-19 outbreak a pandemic and stock market volatility soared \citep{Baker-Bloom-Davis-Kost-Sammon-Viratyosin:2020}. 
This surge is likely due to the BPS approach's tendency to produce spurious rejections when confronted with non-Gaussian data conditions.

\section{Concluding remarks}

This paper has developed a sign-based Monte Carlo resampling method to regularize stock return covariance matrices. Following BPS, the method begins by testing the significance of  pairwise correlations and then sets to zero the sample correlations whose multiplicity-adjusted $p$-values fall above the specified threshold level. A subsequent shrinkage step ensures that the final covariance matrix estimate is positive definite and well-conditioned, while preserving the zero entries achieved by thresholding.

The multiple testing procedures developed in this paper extend the BPS approach by  offering strong control of the traditional FWER, the $k$-FWER,  or the FDP, even in the presence of unknown non-normalities and heteroskedasticity. While the conservative FWER is arguably the most appropriate error rate measure for confirmatory purposes, the more lenient $k$-FWER (with $k>1$) and the FDP may be preferred in exploratory settings where less control is desired in exchange for more power. This was illustrated in an application to portfolio optimization where the goal of multiple testing was not the rigorous validation of candidate hypotheses. Instead, it aimed to regularize stock return covariance matrices, which were then evaluated for their influence on the out-of-sample performance of GMV portfolios.

{\bf Code availability.} An open-access implementation of the procedures developed in this paper is available on GitHub at {https://github.com/richardluger/CovRegMT}.
The R code produces multiplicity-adjusted $p$-values and the associated covariance matrix estimates.

\section*{Acknowledgements}

I wish to thank the editor Serena Ng, an associate editor, and two anonymous reviewers whose constructive feedback greatly improved the final paper.
This work draws on research supported by {l’Autorit{\'e} des march{\'e}s financiers (AMF Qu{\'e}bec)} and the Social Sciences and Humanities Research Council of Canada.

\newpage

\begin{table}[ht]
\begin{center}
\begin{minipage}{6.7in}
\vspace{-0.6cm}
\footnotesize
{\bf Table 1.} 
Empirical rejection rates of FWER procedures
\\[0.5ex]
\begin{tabular*}{\textwidth}{@{\extracolsep{\fill}}  lrrrrrrrrrrrr}
\toprule

			                         &      &	  \multicolumn{3}{c}{Normal}	& 	&   \multicolumn{3}{c}{ $t_{12}$ } 	    & 	&   \multicolumn{3}{c}{ $t_{6}$ } \\

         \cline{3-5} \cline{7-9} \cline{11-13} \\[-2.0ex]
  
			                  &	$T=$	& 	63		&	126	    &	252		&	&  	63		&	126	     &	252	  	&	&  	63	&	126	     &	252	\\

\midrule
 \multicolumn{13}{l}{Panel A: FWER ($\delta=0$) } \\

  \multicolumn{13}{l}{$N=25$ $(M=300)$} \\

 $\text{BPS}_a$			      & 		&	1.2		&	1.9	    &	2.1		&	&	 4.7	&	11.0    &	11.8    &	&	 26.0  	&	 42.3 	&    56.4 \\ 
 $\text{BPS}_b$			      & 		&	2.5		&	3.7	    &	4.4		&	&	10.1	&	18.6    &	20.3    &	&	 38.6  	&	 53.6	&    68.5 \\
 $\text{SS/SD}$               & 		&	5.6		&	5.8     &	4.6		&	&	 6.4	&	 7.0    &	 5.0    &	&	  6.5   &	  5.1   &     4.8 \\[1.0ex]

  \multicolumn{13}{l}{$N=100$ $(M=4950)$} \\

 $\text{BPS}_a$			      & 		&	0.6		&	1.7	    &	3.2		&	&	 5.8	&	14.6    &	23.3    &	&	 29.1  	&	 65.0 	&    87.4 \\
 $\text{BPS}_b$			      & 		&	1.3		&	4.3	    &	5.8		&	&	11.1	&	24.0    &	36.4    &	&	 42.8  	&	 77.7	&    93.7 \\
 $\text{SS/SD}$               & 		&	5.8		&	6.6     &	6.0		&	&	 7.6	&	 5.2    &	 5.1    &	&	  6.4   &	  5.1   &     7.2 \\[1.0ex]

  \multicolumn{13}{l}{$N=500$ $(M=124750)$} \\

 $\text{BPS}_a$			      &	  	    &	0.0		&	0.8	    &	1.9		&	&	  2.3  	&	 21.0	&	47.1    &	&	  32.2	&	  85.7	&	 98.4	\\    
 $\text{BPS}_b$			      &		    &	0.3		&	1.6	    &	4.4		&	&	  4.9  	&	 33.1 	&	61.0    &	&	  45.7 	&	  93.3 	&	 99.5 	\\
 $\text{SS/SD}$			      & 		&	6.8		&	5.4	    &	4.8		&	&     5.5	&	  5.1   &    5.8    &	&	   7.4  &	   5.8  &	  4.5	\\

\midrule
 \multicolumn{13}{l}{Panel B: FWER ($\delta=0.9$) } \\

  \multicolumn{13}{l}{$N=25$ $(M=300)$} \\

 $\text{BPS}_a$               &         &    0.4    &   0.6     &   0.5     &   &   1.4     &    2.1    &      2.7      &   &     5.9   &    9.7    &    14.5   \\ 
 $\text{BPS}_b$               &         &    0.6    &   1.1     &   0.7     &   &   2.2     &    3.3    &      4.5      &   &     8.9   &   14.4   &     20.7    \\
 $\text{SS}$                  &         &    0.0    &   0.0     &   0.0     &   &   0.2     &    0.0    &      0.0      &   &     0.2   &    0.0    &     0.0   \\
 $\text{SD}$                  &         &    0.9    &   1.9     &   1.9     &   &   0.9     &    2.0    &      2.1      &   &     1.4   &    1.2    &     2.2   \\[1.0ex]

  \multicolumn{13}{l}{$N=100$ $(M=4950)$} \\

 $\text{BPS}_a$               &         &    0.0    &   0.2     &   0.4     &   &   0.7     &    2.9    &      3.6      &   &     7.0   &   19.0    &    30.0   \\ 
 $\text{BPS}_b$               &         &    0.0    &   0.4     &   1.0     &   &   1.4     &    4.1    &      5.7      &   &    10.1   &   25.9    &    37.6    \\
 $\text{SS}$                  &         &    0.0    &   0.0     &   0.0     &   &   0.0     &    0.0    &      0.0      &   &     0.0   &    0.1    &     0.1   \\
 $\text{SD}$                  &         &    0.2    &   0.8     &   1.9     &   &   0.2     &    0.9    &      0.9      &   &     0.7   &    0.8    &     1.1   \\[1.0ex]

  \multicolumn{13}{l}{$N=500$ $(M=124750)$} \\
    
 $\text{BPS}_a$               &         &    0.0    &   0.5     &   0.6     &   &    0.4    &     3.4   &       7.9     &   &     8.3   &   38.0    &     68.3  \\    
 $\text{BPS}_b$               &         &    0.1    &   0.8     &   0.9     &   &    0.7    &     5.4   &      12.9     &   &    14.0   &   47.8    &     76.4  \\
 $\text{SS}$                  &         &    0.0    &   0.0     &   0.0     &   &    0.0    &     0.0   &       0.0     &   &     0.1   &    0.0    &      0.0  \\
 $\text{SD}$                  &         &    0.2    &   0.7     &   0.8     &   &    0.1    &     0.9   &       1.1     &   &     0.3   &    1.2    &      1.6  \\

\midrule

 \multicolumn{13}{l}{Panel C: Average power ($\delta=0.9$)} \\

  \multicolumn{13}{l}{$N=25$ $(M=300)$} \\

 $\text{BPS}_a/\text{BPS}_b$  &         &    53.8   &   69.8    &   80.6    &   &    48.9   &    64.4   &     76.9      &   &    40.7   &    57.8   &   69.1    \\
 $\text{SS}$                  &         &    32.5   &   50.4    &   66.0    &   &    30.1   &    47.0   &     62.1      &   &    26.1   &    41.3   &   55.5    \\
 $\text{SD}$                  &         &    45.5   &   64.6    &   78.6    &   &    41.9   &    60.4   &     74.5      &   &    36.3   &    53.1   &   67.4    \\[1.0ex]

  \multicolumn{13}{l}{$N=100$ $(M=4950)$} \\

 $\text{BPS}_a/\text{BPS}_b$  &         &    43.2   &   61.4    &   75.3    &   &    38.9   &    56.1   &     70.6      &   &    30.5   &    45.5   &   59.9   \\
 $\text{SS}$                  &         &    21.3   &   38.9    &   56.0    &   &    19.5   &    35.6   &     52.2      &   &    16.8   &    30.7   &   45.9    \\
 $\text{SD}$                  &         &    34.4   &   55.7    &   72.2    &   &    30.8   &    50.5   &     67.4      &   &    25.7   &    42.5   &   58.4   \\[1.0ex]

  \multicolumn{13}{l}{$N=500$ $(M=124750)$} \\
    
 $\text{BPS}_a/\text{BPS}_b$  &         &    33.7   &   53.2    &   69.6    &   &    29.1   &    46.5   &     62.4      &   &    19.5   &   33.0    &   44.9    \\
 $\text{SS}$                  &         &    12.8   &   28.8    &   46.8    &   &    11.2   &    25.7   &     42.6      &   &     9.6   &   21.8    &   36.0    \\
 $\text{SD}$                  &         &    23.8   &   46.0    &   65.3    &   &    20.7   &    40.3   &     58.8      &   &    16.6   &   32.3    &   47.3    \\

\bottomrule

\end{tabular*}
\\[1.0ex]
Notes: Panels A and B report the empirical FWER, while Panel C  reports the empirical average power, in percentages, given a nominal FWER of $5\%$. The power results for the $\text{BPS}$ procedures are based on strong FWER-adjusted critical values. 

\end{minipage}
\end{center}
\end{table}

\newpage

\begin{table}[ht]
\begin{center}
\begin{minipage}{6.7in}
\vspace{-0.6cm}
\footnotesize
{\bf Table 2.} 
Empirical rejection rates of $k$-FWER procedures when $\delta=0.9$
\\[0.5ex]
\begin{tabular*}{\textwidth}{@{\extracolsep{\fill}}  lrrrrrrrrrrrr}
\toprule

			            &      &	  \multicolumn{3}{c}{Normal}	& 	&   \multicolumn{3}{c}{ $t_{12}$ } 	    & 	&   \multicolumn{3}{c}{ $t_{6}$ } \\

         \cline{3-5} \cline{7-9} \cline{11-13} \\[-2.0ex]
  
			            &	$T=$	& 	63	  &	126       &	252		&	&  	63		&	126	      &	252	  		  &	 &  	63	&	126	     &	252	\\

\midrule
 \multicolumn{13}{l}{Panel A: $k$-FWER  } \\[1.0ex]

  \multicolumn{13}{l}{$N=25$ $(M=300)$} \\

$\text{SS}_{5}$  					&         &    0.0    &   0.1     &   0.1     &   &   0.3     &    0.1    &      0.3      &   &     0.3   &    0.0    &     0.0   \\
$\text{SD}_{5}$ 			 		&         &    0.1    &   0.2     &   0.3     &   &   0.3     &    0.3    &      0.5      &   &     0.3   &    0.0    &     0.1   \\
$\text{SS}_{17}$ 					&         &    0.0    &   0.0     &   0.0     &   &   0.0     &    0.1    &      0.0      &   &     0.0   &    0.0    &     0.0   \\
$\text{SD}_{17}$ 					&         &    0.0    &   0.0     &   0.0     &   &   0.0     &    0.1    &      0.0      &   &     0.0   &    0.0    &     0.0   \\[1.0ex]

  \multicolumn{13}{l}{$N=100$ $(M=4950)$} \\

$\text{SS}_{8}$  					&         &    0.0    &   0.0     &   0.0     &   &   0.0     &    0.0    &      0.0      &   &     0.1   &    0.0    &     0.0   \\
$\text{SD}_{8}$  					&         &    0.0    &   0.1     &   0.0     &   &   0.0     &    0.0    &      0.0      &   &     0.2   &    0.0    &     0.0   \\
$\text{SS}_{70}/\text{SD}_{70}$ 	&         &    0.0    &   0.0     &   0.0     &   &   0.0     &    0.0    &      0.0      &   &     0.0   &    0.0    &     0.0   \\[1.0ex]

  \multicolumn{13}{l}{$N=500$ $(M=124750)$} \\
    
$\text{SS}_{11}$  					&         &    0.0    &   0.0     &   0.0     &   &   0.0     &    0.0    &      0.0      &   &     0.0   &    0.0    &     0.0   \\
$\text{SD}_{11}$			  		&         &    0.0    &   0.0     &   0.1     &   &   0.0     &    0.0    &      0.0      &   &     0.0   &    0.0    &     0.0   \\
$\text{SS}_{353}/\text{SD}_{353}$ 	&         &    0.0    &   0.0     &   0.0     &   &   0.0     &    0.0    &      0.0      &   &     0.0   &    0.0    &     0.0   \\[1.0ex]

\midrule

 \multicolumn{13}{l}{Panel B: Average power } \\[1.0ex]

  \multicolumn{13}{l}{$N=25$ $(M=300)$} \\

$\text{SS}_{5}$  					&         &    54.4   &   69.2    &   80.5    &   &   51.5    &    65.8   &      77.4     &   &     46.6  &    60.6   &     71.9   \\
$\text{SD}_{5}$  					&         &    55.8   &   70.9    &   82.0    &   &   52.8    &    67.5   &      79.1     &   &     47.7  &    62.0   &     73.5  \\
$\text{SS}_{17}$ 					&         &    67.6   &   79.3    &   87.4    &   &   65.1    &    76.5   &      85.3     &   &     60.5  &    72.1   &     81.0  \\
$\text{SD}_{17}$ 					&         &    67.7   &   79.4    &   87.5    &   &   65.2    &    76.6   &      85.4     &   &     60.6  &    72.2   &     81.2 \\[1.0ex]

  \multicolumn{13}{l}{$N=100$ $(M=4950)$} \\

$\text{SS}_{8}$ 			 		&         &    40.6   &   57.6    &   71.6    &   &   37.8    &    54.2   &      68.1     &   &     33.6  &    48.3   &     62.3  \\
$\text{SD}_{8}$  					&         &    41.9   &   59.8    &   74.0    &   &   38.8    &    55.9   &      70.3     &   &     34.3  &    49.6   &     63.9 \\
$\text{SS}_{70}/\text{SD}_{70}$ 	&         &    58.3   &   72.2    &   82.3    &   &   55.3    &    69.1   &      79.6     &   &     50.7  &    63.8   &     75.0 \\[1.0ex]

  \multicolumn{13}{l}{$N=500$ $(M=124750)$} \\
    
$\text{SS}_{11}$  					&         &    27.0   &   44.9    &   61.3    &   &   24.5    &    41.1   &      57.3     &   &     21.3  &    36.0   &     50.4  \\
$\text{SD}_{11}$  					&         &    28.2   &   47.6    &   65.6    &   &   25.4    &    43.1   &      60.3     &   &     21.9  &    37.1   &     51.9 \\
$\text{SS}_{353}/\text{SD}_{353}$ 	&         &    47.8   &   63.7    &   76.1    &   &   44.6    &    60.0   &      72.9     &   &     39.9  &    54.4   &     66.8 \\[1.0ex]

\bottomrule

\end{tabular*}
\\[1.0ex]
Notes: Panel A reports the empirical $k$-FWER and Panel B  reports the empirical average power, in percentages, given a nominal $k$-FWER of  $5\%$. 
The $\text{SS}_{k}$ and $\text{SD}_{k}$ results are based on the rules $k=\lfloor \log{M} \rfloor$ and $k=\lfloor \sqrt{M} \rfloor$.
When the value of $k$ is large, the $\text{SS}_{k}$ and $\text{SD}_{k}$ results converge, and in those cases the table presents the results on the same line.

\end{minipage}
\end{center}
\end{table}

\newpage

\begin{table}[ht]
\begin{center}
\begin{minipage}{6.7in}
\vspace{-0.6cm}
\footnotesize
{\bf Table 3.} 
Empirical  exceedance probability and average power  of FDP procedures when $\delta=0.9$
\\[0.5ex]
\begin{tabular*}{\textwidth}{@{\extracolsep{\fill}}  lrrrrrrrrrrrr}
\toprule

			            &      &	  \multicolumn{3}{c}{Normal}	& 	&   \multicolumn{3}{c}{ $t_{12}$ } 	    & 	&   \multicolumn{3}{c}{ $t_{6}$ } \\

         \cline{3-5} \cline{7-9} \cline{11-13} \\[-2.0ex]
  
			            			&	$T=$   & 	63	  &	126       &	252		  &	 &  	63		&	126	      &	252	  		  &	 &  	63	&	126	     &	252	\\

\midrule
 \multicolumn{13}{l}{Panel A: Exceedance probability  } \\[1.0ex]

  \multicolumn{13}{l}{$N=25$ $(M=300)$} \\

$\text{SS}_{0.1}$ 					&         &    0.0    &   0.0     &   0.0     &   &   0.0     &    0.1    &      0.0      &   &     0.0   &    0.0    &     0.0   \\
$\text{SD}_{0.1}$ 					&         &    0.0    &   0.0     &   0.0     &   &   0.0     &    0.1    &      0.0      &   &     0.0   &    0.0    &     0.0   \\[1.0ex]

  \multicolumn{13}{l}{$N=100$ $(M=4950)$} \\

$\text{SS}_{0.1}/\text{SD}_{0.1}$ 	&         &    0.0    &   0.0     &   0.0     &   &   0.0     &    0.0    &      0.0      &   &     0.0   &    0.0    &     0.0   \\[1.0ex]

  \multicolumn{13}{l}{$N=500$ $(M=124750)$} \\
    
$\text{SS}_{0.1}/\text{SD}_{0.1}$ 	&         &    0.0    &   0.0     &   0.0     &   &   0.0     &    0.0    &      0.0      &   &     0.0   &    0.0    &     0.0   \\[1.0ex]

\midrule

 \multicolumn{13}{l}{Panel B: Average power } \\[1.0ex]

  \multicolumn{13}{l}{$N=25$ $(M=300)$} \\

$\text{SS}_{0.1}$ 					&         &    66.0   &   79.5    &   88.1    &   &   62.9    &    76.5   &      85.9    &   &     57.0  &    71.4   &     81.4 \\
$\text{SD}_{0.1}$ 					&         &    66.1   &   79.6    &   88.1    &   &   63.0    &    76.6   &      86.0    &   &     57.1  &    71.5   &     81.5 \\[1.0ex]

  \multicolumn{13}{l}{$N=100$ $(M=4950)$} \\

$\text{SS}_{0.1}/\text{SD}_{0.1}$	&         &    70.4   &   82.0    &   89.3    &   &   67.4    &    79.5   &      87.4     &   &     62.3 &    74.7   &     83.9  \\[1.0ex]

  \multicolumn{13}{l}{$N=500$ $(M=124750)$} \\

$\text{SS}_{0.1}/\text{SD}_{0.1}$	&         &    71.2   &   82.4    &   89.5    &   &   68.1    &    79.8   &      87.6     &   &     62.8 &    75.2   &     83.8  \\[1.0ex]

\bottomrule

\end{tabular*}
\\[1.0ex]
Notes: Panel A reports the empirical estimate of $\Pr \left(   \text{FDP} > 0.1   \right)$ and Panel B  reports the empirical average 
power, in percentages, given the nominal error criterion $\Pr \left(   \text{FDP} > 0.1   \right) \le 5\%$. 
When the value of $M$ is large, the $\text{SS}_{0.1}$ and $\text{SD}_{0.1}$ results converge, and in those cases the table presents the results on the same line.

\end{minipage}
\end{center}
\end{table}

\newpage

\newpage
\clearpage

\thispagestyle{empty}

\newpage

\thispagestyle{empty}

\begin{landscape}

\begin{table}
\begin{center}
\hspace*{-35pt}
\begin{minipage}{10.1in}
\footnotesize
{\bf Table 4.} Out-of-sample performance of the portfolio strategies when $N=100$ $(M=4950)$ 
\\[0.5ex]
\begin{tabular*}{\textwidth}{@{\extracolsep{\fill}}  lcllllllcllllllcllllll}
\toprule

                   						          & &      \multicolumn{6}{c}{$L=63$}                           & & \multicolumn{6}{c}{$L=126$}                            & &   \multicolumn{6}{c}{$L=252$}     \\

                                    \cline{3-8} \cline{10-15} \cline{17-22} \\[-2.0ex]
  
                   						          & &   AV   &   SD            &   IR     &   TO   &  MDD     &  TW     & &  AV     &    SD        &  IR    &    TO    & MDD   &  TW   & &  AV   &  SD         & IR   &   TO   & MDD    &  TW  \\
\midrule 
  \multicolumn{22}{l}{Panel A: Short selling allowed} \\
 
Sample             						          & &    n/a &    n/a          &   n/a    &   n/a  &  n/a     &  n/a   & &    -5.25 &  $22.49^{***}$ &  -0.23 &  10.54   & 82.57 &  0.26 & &  6.33 & $14.32^{*}$   & 0.44 &   2.86 & 48.78 & 2.45 \\
EW                 						          & &  16.60 &   $19.23^{***}$ &   0.86   &   0.10 &  43.43   &  12.12 & &    16.63 &  $19.23^{***}$ &  0.86  &  0.10    & 43.42 & 12.17 & & 16.64 & $19.23^{***}$ & 0.86 &   0.10 & 43.56 & 12.20 \\
VT                 						          & &  14.86 &   $16.54$       &   0.89   &   0.22 &  36.90   &   9.79 & &    14.79 &  $16.65^{***}$ &  0.88  &  0.15    & 37.44 &  9.65 & & 14.77 & $16.75^{***}$ & 0.88 &   0.11 & 38.15 &  9.60 \\
$\text{BPS}_a$     						          & &  13.18 &   $16.25$       &   0.81   &   1.56 &  42.51   &   7.43 & &    13.62 &  $15.24$       &  0.89  &  1.50    & 31.99 &  8.23 & & 11.36 & $15.05$       & 0.75 &   1.44 & 33.94 &  5.64 \\
$\text{BPS}_b$     						          & &  13.04 &   $\bf 16.31$   &   0.79   &   1.68 &  41.82   &   7.26 & &    13.24 &  $\bf 15.38$   &  0.86  &  1.57    & 31.06 &  7.69 & & 11.34 & $\bf 15.04$   & 0.75 &   1.51 & 33.28 &  5.62 \\
SS                 						          & &  15.14 &   $16.52$       &   0.91   &   1.10 &  33.95   &  10.28 & &    16.06 &  $15.74$       &  1.02  &  1.08    & 29.83 & 12.27 & & 14.90 & $15.36$       & 0.97 &   0.88 & 30.25 & 10.18 \\
SD                 						          & &  14.28 &   $16.43$       &   0.86   &   1.40 &  40.29   &   8.91 & &    14.17 &  $15.71$       &  0.90  &  1.48    & 32.27 &  8.92 & & 13.56 & $14.95$       & 0.90 &   1.40 & 33.11 &  8.20 \\
$\text{SS}_{8}$   								  & &  14.39 &   $16.29$       &   0.88   &   1.81 &  38.92   &   9.11 & &   14.68  &  $15.54$       &  0.94  &  1.49    & 30.11 &  9.76 & & 12.91 & $14.93$       & 0.86 &   1.19 & 32.14 &  7.36 \\
$\text{SD}_{8}$   								  & &  14.31 &   $16.38$       &   0.87   &   1.82 &  38.38   &   8.98 & &   13.72  &  $15.57$       &  0.88  &  1.64    & 35.25 &  8.29 & & 13.08 & $14.87$       & 0.87 &   1.48 & 37.36 &  7.57 \\
$\text{SS}_{70}$  								  & &  13.87 &   $15.75^{***}$ &   0.88   &   2.42 &  33.72   &   8.46 & &   14.03  &  $15.49$       &  0.90  &  2.03    & 39.19 &  8.76 & & 13.44 & $14.73^{**}$  & 0.91 &   1.82 & 34.08 &  8.08 \\
$\text{SD}_{70}$  								  & &  13.90 &   $15.75^{**}$  &   0.88   &   2.42 &  32.97   &   8.50 & &   13.32  &  $15.50$       &  0.85  &  2.08    & 40.36 &  7.77 & & 13.72 & $14.68^{**}$  & 0.93 &   1.89 & 31.77 &  8.49 \\
$\text{SS}_{0.1}$                    			  & &  12.43 &   $17.42^{*}$   &   0.71   &   3.17 &  33.75   &   6.35 & &    8.83  &  $15.40$       &  0.57  &  2.85    & 39.01 &  3.64 & & 11.83 & $14.97$       & 0.79 &   2.53 & 36.16 &  6.12 \\
$\text{SD}_{0.1}$                    		      & &  12.31 &   $17.36^{*}$   &   0.70   &   3.17 &  33.45   &   6.22 & &    9.65  &  $15.43$       &  0.62  &  2.87    & 36.97 &  4.18 & & 11.91 & $14.92$       & 0.79 &   2.51 & 37.50 &  6.21 \\
LS                 						          & &  10.29 &   $13.22^{***}$ &   0.77   &   2.44 &  39.60   &   4.91 & &     9.97 &  $13.24^{***}$ &  0.75  &  2.22    & 31.64 &  4.66 & &  8.71 & $13.02^{***}$ & 0.66 &   1.60 & 38.67 &  3.77 \\
NLS                						          & &  11.59 &   $12.77^{***}$ &   0.90   &   1.70 &  36.00   &   6.19 & &    10.93 &  $12.86^{***}$ &  0.84  &  1.99    & 31.58 &  5.52 & &  9.33 & $12.75^{***}$ & 0.73 &   1.53 & 37.92 &  4.22 \\[1.0ex]

  \multicolumn{22}{l}{Panel B: Short selling prohibited} \\

Sample             						          & &    n/a &   n/a           &   n/a    &   n/a  &   n/a    &    n/a  & &   11.61 &  $13.20^{***}$ &  0.87  &   0.63   & 31.62 &  6.15 & & 10.48 & $13.22^{***}$ & 0.79 &   0.40 & 31.81 & 5.08  \\
EW                 						          & &  16.60 &   $19.23^{***}$ &   0.86   &   0.10 &  43.43   &  12.12  & &   16.63 &  $19.23^{***}$ &  0.86  &   0.10   & 43.42 & 12.17 & & 16.64 & $19.23^{***}$ & 0.86 &   0.10 & 43.56 & 12.20  \\
VT                 						          & &  14.86 &   $16.54^{***}$ &   0.89   &   0.22 &  36.90   &   9.79  & &   14.79 &  $16.65^{***}$ &  0.88  &   0.15   & 37.44 &  9.65 & & 14.77 & $16.75^{***}$ & 0.88 &   0.11 & 38.15 &  9.60 \\
$\text{BPS}_a$     						          & &  14.41 &   $14.99^{***}$ &   0.96   &   0.46 &  37.21   &   9.46  & &   13.94 &  $14.44$       &  0.96  &   0.40   & 33.72 &  8.86 & & 12.92 & $14.01^{**}$  & 0.92 &   0.34 & 33.41 &  7.54 \\
$\text{BPS}_b$     						          & &  14.37 &   $\bf 14.90$   &   0.96   &   0.48 &  37.46   &   9.42  & &   13.77 &  $\bf 14.41$   &  0.95  &   0.42   & 33.65 &  8.61 & & 12.81 & $\bf 13.97$   & 0.91 &   0.35 & 33.35 &  7.40 \\
SS                 						          & &  14.82 &   $15.70^{***}$ &   0.94   &   0.38 &  35.83   &   9.96  & &   14.66 &  $15.48^{***}$ &  0.94  &   0.33   & 34.21 &  9.75 & & 14.54 & $15.25^{***}$ & 0.95 &   0.29 & 33.85 &  9.61  \\
SD                 						          & &  14.42 &   $15.38^{***}$ &   0.93   &   0.44 &  37.49   &   9.38  & &   13.60 &  $14.97^{***}$ &  0.90  &   0.39   & 34.37 &  8.25 & & 13.19 & $14.50^{***}$ & 0.90 &   0.34 & 34.16 &  7.79 \\
$\text{SS}_{8}$   							      & &  14.37 &   $15.02$       &   0.95   &   0.51 &  35.63   &   9.39  & &   13.99 &  $14.80^{***}$ &  0.94  &   0.41   & 34.25 &  8.85 & & 13.54 & $14.42^{***}$ & 0.93 &   0.33 & 33.25 &  8.28  \\
$\text{SD}_{8}$   								  & &  14.29 &   $15.00$       &   0.95   &   0.52 &  36.24   &   9.27  & &   13.59 &  $14.73^{***}$ &  0.92  &   0.43   & 34.43 &  8.29 & & 13.17 & $14.24^{***}$ & 0.92 &   0.35 & 33.07 &  7.82 \\
$\text{SS}_{70}$					     		  & &  14.59 &   $14.28^{***}$ &   1.02   &   0.61 &  32.42   &   9.93  & &   13.62 &  $14.16^{***}$ &  0.96  &   0.49   & 31.21 &  8.45 & & 12.83 & $13.70^{***}$ & 0.93 &   0.38 & 31.67 &  7.48  \\
$\text{SD}_{70}$  		  					      & &  14.61 &   $14.28^{***}$ &   1.02   &   0.61 &  32.15   &   9.95  & &   13.54 &  $14.16^{***}$ &  0.95  &   0.49   & 31.45 &  8.33 & & 12.86 & $13.68^{***}$ & 0.94 &   0.38 & 31.46 &  7.52  \\
$\text{SS}_{0.1}$			                      & &  14.01 &   $13.74^{***}$ &   1.01   &   0.71 &  31.39   &   9.11  & &   13.04 &  $13.60^{***}$ &  0.95  &   0.56   & 31.11 &  7.76 & & 11.79 & $13.53^{***}$ & 0.87 &   0.40 & 31.50 &  6.29 \\
$\text{SD}_{0.1}$             				      & &  14.01 &   $13.75^{***}$ &   1.01   &   0.71 &  31.39   &   9.11  & &   13.06 &  $13.61^{***}$ &  0.95  &   0.56   & 31.10 &  7.78 & & 11.80 & $13.53^{***}$ & 0.87 &   0.40 & 31.73 &  6.30 \\
LS                 						          & &  13.71 &   $13.19^{***}$ &   1.03   &   0.80 &  32.38   &   8.78  & &   12.27 &  $13.17^{***}$ &  0.93  &   0.55   & 32.10 &  6.88 & & 10.92 & $13.21^{***}$ & 0.82 &   0.36 & 31.87 &  5.47 \\
NLS                						          & &  13.52 &   $13.15^{***}$ &   1.02   &   0.72 &  32.74   &   8.50  & &   12.48 &  $13.10^{***}$ &  0.95  &   0.50   & 31.21 &  7.14 & & 11.25 & $13.22^{***}$ & 0.85 &   0.35 & 31.66 &  5.79 \\

\bottomrule

\end{tabular*}
\\[1.0ex]
{\bf Notes:} This table reports the  
annualized mean (AV, in \%), 
annualized standard deviation (SD, in \%), 
 information ratio (IR),
 average turnover (TO),
maximum drawdown (MDD, in \%),
and terminal wealth (TW, in \$)
  achieved by the rolling-window  portfolio strategies with monthly rebalancing, given an initial wealth of \$1. 
The trading period is from January 3, 2005 to  December 31, 2021, and
$L$ is the length (in days) of the estimation window. 
One, two, and three stars indicate that SD is statistically different from  SD of the benchmark $\text{BPS}_b$ strategy (in bold)
at the 10\%, 5\%, and 1\% levels, respectively.

\end{minipage}
\end{center}
\end{table}

\end{landscape}

\newpage

\thispagestyle{empty}

\begin{landscape}

\begin{table}
\begin{center}
\hspace*{-35pt}
\begin{minipage}{10.1in}
\footnotesize
{\bf Table 5.} Out-of-sample performance of the portfolio strategies when $N=500$ $(M=124750)$
\\[0.5ex]
\begin{tabular*}{\textwidth}{@{\extracolsep{\fill}}  lcllllllcllllllcllllll}
\toprule

                   						& &      \multicolumn{6}{c}{$L=63$}                           & & \multicolumn{6}{c}{$L=126$}                            & &   \multicolumn{6}{c}{$L=252$}     \\

                                    \cline{3-8} \cline{10-15} \cline{17-22} \\[-2.0ex]
  
                   						          & &   AV   &   SD            &   IR    &   TO     &  MDD     &  TW     & &   AV     &    SD           &  IR    &    TO   & MDD   &  TW   & &  AV    &  SD            & IR    & TO    & MDD     &  TW  \\
\midrule 
  \multicolumn{22}{l}{Panel A: Short selling allowed} \\
 
Sample             						          & &  n/a   &   n/a           &   n/a    &   n/a    &  n/a     & n/a     & &   n/a    &   n/a          &  n/a   &  n/a    & n/a   & n/a    & &   n/a & n/a            & n/a   &  n/a  & n/a   & n/a    \\
EW                 						          & &  20.29 &   $20.47^{***}$ &   0.99   &   0.10   &  44.94   & 21.68   & &   20.33  &  $20.48^{***}$ &  0.99  &  0.10   & 44.99 &  21.82 & & 20.25 &  $20.52^{***}$ &  0.98 &  0.10 & 44.87 & 21.51 \\
VT                 						          & &  16.63 &   $16.26^{***}$ &   1.02   &   0.27   &  37.22   & 13.30   & &   16.88  &  $17.08^{***}$ &  0.98  &  0.17   & 38.28 &  13.58 & & 17.35 &  $17.60^{***}$ &  0.98 &  0.10 & 38.43 & 14.48 \\
$\text{BPS}_a$     						          & &  16.69 &   $15.63^{*}$   &   1.06   &   1.07   &  37.81   & 13.67   & &   16.37  &  $15.56^{**}$  &  1.05  &  1.23   & 37.65 &  12.98 & & 16.40 &  $14.65$       &  1.11 &  1.12 & 29.31 & 13.37 \\
$\text{BPS}_b$     						          & &  16.62 &   $\bf 15.58$   &   1.06   &   1.12   &  38.76   & 13.53   & &   16.43  &  $\bf 15.22$   &  1.07  &  1.24   & 35.23 &  13.24 & & 16.39 &  $\bf 14.62$   &  1.12 &  1.16 & 29.23 & 13.35 \\
SS                 						          & &  17.03 &   $16.53^{***}$ &   1.03   &   0.68   &  36.01   & 14.15   & &   17.10  &  $16.99^{***}$ &  1.00  &  0.89   & 34.93 &  14.12 & & 18.09 &  $17.10^{***}$ &  1.05 &  0.87 & 34.43 & 16.63 \\
SD                 						          & &  16.71 &   $16.41^{***}$ &   1.01   &   0.80   &  36.05   & 13.43   & &   16.45  &  $16.60^{***}$ &  0.99  &  1.01   & 35.96 &  12.80 & & 17.27 &  $16.24^{***}$ &  1.06 &  1.01 & 33.65 & 14.85 \\
$\text{SS}_{11}$								  & &  16.42 &   $16.10^{*}$   &   1.01   &   1.18   &  36.59   & 12.90   & &   16.15  &  $16.09^{***}$ &  1.00  &  1.13   & 36.25 &  12.33 & & 17.08 &  $15.76^{***}$ &  1.08 &  0.92 & 34.20 & 14.56 \\
$\text{SD}_{11}$							      & &  16.37 &   $16.05^{*}$   &   1.01   &   1.18   &  36.58   & 12.81   & &   16.16  &  $16.06^{***}$ &  1.00  &  1.14   & 36.08 &  12.36 & & 17.09 &  $15.66^{***}$ &  1.09 &  0.93 & 34.27 & 14.62 \\
$\text{SS}_{353}$     	 						  & &  15.82 &   $14.69^{***}$ &   1.07   &   1.69   &  35.39   & 12.09   & &   17.36  &  $14.79^{**}$  &  1.17  &  1.55   & 35.89 &  15.66 & & 18.40 &  $14.63$       &  1.25 &  1.40 & 30.87 & 18.74 \\
$\text{SD}_{353}$  								  & &  15.82 &   $14.69^{***}$ &   1.07   &   1.69   &  35.39   & 12.09   & &   17.36  &  $14.79^{**}$  &  1.17  &  1.55   & 35.89 &  15.65 & & 18.38 &  $14.63$       &  1.25 &  1.40 & 30.87 & 18.69 \\
$\text{SS}_{0.1}$                    			  & &  14.45 &   $13.45^{***}$ &   1.07   &   2.57   &  35.10   &  9.90   & &   17.83  &  $14.02^{***}$ &  1.27  &  2.85   & 29.35 &  17.28 & & 16.88 &  $15.51^{**}$  &  1.08 &  3.54 & 34.49 & 14.19 \\
$\text{SD}_{0.1}$                    			  & &  14.45 &   $13.45^{***}$ &   1.07   &   2.57   &  35.10   &  9.90   & &   17.83  &  $14.02^{***}$ &  1.27  &  2.85   & 29.35 &  17.28 & & 16.88 &  $15.51^{**}$  &  1.08 &  3.54 & 34.49 & 14.19 \\
LS                 						          & &  12.57 &   $11.45^{***}$ &   1.09   &   2.04   &  33.56   &  7.50   & &   10.57  &  $11.22^{***}$ &  0.94  &  2.34   & 32.24 &   5.37 & &  7.98 &  $11.19^{***}$ &  0.71 &  2.96 & 36.26 &  3.47 \\
NLS                						          & &  12.49 &   $11.39^{***}$ &   1.09   &   1.67   &  34.24   &  7.41   & &   11.16  &  $10.99^{***}$ &  1.01  &  1.57   & 32.46 &   5.96 & &  8.81 &  $10.57^{***}$ &  0.83 &  1.51 & 37.57 &  4.03 \\[1.0ex]

  \multicolumn{22}{l}{Panel B: Short selling prohibited} \\

Sample             						          & & n/a    &   n/a           &   n/a    &   n/a    &   n/a    &    n/a  & &   n/a    &   n/a          &  n/a   &   n/a   &  n/a  & n/a   & & n/a   & n/a            & n/a   &  n/a  & n/a   & n/a       \\
EW                 						          & & 20.29  &   $20.47^{***}$ &   0.99   &   0.10   &  44.94   &  21.68  & &   20.33  &  $20.48^{***}$ &  0.99  &   0.10  & 44.99 & 21.82 & & 20.25 &  $20.52^{***}$ &  0.98 &  0.10 & 44.87 & 21.51   \\
VT                 						          & & 16.63  &   $16.26^{***}$ &   1.02   &   0.27   &  37.22   &  13.30  & &   16.88  &  $17.08^{***}$ &  0.98  &   0.17  & 38.28 & 13.58 & & 17.35 &  $17.60^{***}$ &  0.98 &  0.10 & 38.43 & 14.48 \\
$\text{BPS}_a$     						          & & 15.86  &   $14.65^{***}$ &   1.08   &   0.42   &  36.63   &  12.20  & &   15.66  &  $14.44^{***}$ &  1.08  &   0.35  & 34.59 & 11.84 & & 15.31 &  $14.28^{***}$ &  1.07 &  0.27 & 31.89 & 11.22 \\
$\text{BPS}_b$     						          & & 15.78  &   $\bf 14.52$   &   1.08   &   0.43   &  36.60   &  12.07  & &   15.59  &  $\bf 14.33$   &  1.08  &   0.36  & 34.39 & 11.74 & & 15.22 &  $\bf 14.19$   &  1.07 &  0.27 & 31.58 & 11.06 \\
SS                 						          & & 16.21  &   $15.63^{***}$ &   1.03   &   0.35   &  36.61   &  12.61  & &   16.30  &  $16.13^{***}$ &  1.01  &   0.28  & 37.46 & 12.63 & & 16.85 &  $16.25^{***}$ &  1.03 &  0.21 & 36.62 & 13.83  \\
SD                 						          & & 16.01  &   $15.38^{***}$ &   1.04   &   0.37   &  36.89   &  12.28  & &   15.88  &  $15.72^{***}$ &  1.01  &   0.31  & 37.35 & 11.90 & & 16.13 &  $15.67^{***}$ &  1.02 &  0.24 & 36.40 & 12.43  \\
$\text{SS}_{11}$							      & & 15.62  &   $14.78^{**}$  &   1.05   &   0.44   &  36.40   &  11.67  & &   15.73  &  $15.20^{***}$ &  1.03  &   0.33  & 36.75 & 11.76 & & 16.16 &  $15.29^{***}$ &  1.05 &  0.24 & 35.45 & 12.62 \\
$\text{SD}_{11}$ 			   				      & & 15.62  &   $14.77^{**}$  &   1.05   &   0.44   &  36.40   &  11.67  & &   15.71  &  $15.17^{***}$ &  1.03  &   0.34  & 36.73 & 11.72 & & 16.09 &  $15.23^{***}$ &  1.05 &  0.24 & 35.34 & 12.49 \\
$\text{SS}_{353}$  								  & & 14.71  &   $13.37^{***}$ &   1.09   &   0.57   &  34.18   &  10.34  & &   15.01  &  $13.86^{***}$ &  1.08  &   0.43  & 33.54 & 10.76 & & 15.45 &  $14.00^{*}$   &  1.10 &  0.30 & 31.31 & 11.56 \\
$\text{SD}_{353}$  								  & & 14.71  &   $13.37^{***}$ &   1.09   &   0.57   &  34.18   &  10.34  & &   15.01  &  $13.86^{***}$ &  1.08  &   0.43  & 33.54 & 10.76 & & 15.45 &  $14.00^{*}$   &  1.10 &  0.30 & 31.31 & 11.56 \\
$\text{SS}_{0.1}$                    			  & & 12.92  &   $11.42^{***}$ &   1.13   &   0.72   &  30.55   &   7.97  & &   12.80  &  $11.88^{***}$ &  1.07  &   0.58  & 29.34 &  7.73 & & 13.71 &  $12.39^{***}$ &  1.10 &  0.41 & 26.24 &  8.93 \\
$\text{SD}_{0.1}$                    			  & & 12.92  &   $11.42^{***}$ &   1.13   &   0.72   &  30.55   &   7.97  & &   12.80  &  $11.88^{***}$ &  1.07  &   0.58  & 29.34 &  7.73 & & 13.71 &  $12.39^{***}$ &  1.10 &  0.41 & 26.24 &  8.93  \\
LS                 						          & & 13.70  &   $11.35^{***}$ &   1.20   &   0.99   &  30.25   &   9.10  & &   13.24  &  $11.51^{***}$ &  1.15  &   0.73  & 26.49 &  8.39 & & 13.19 &  $11.80^{***}$ &  1.11 &  0.50 & 22.90 &  8.27 \\
NLS                						          & & 15.09  &   $11.67^{***}$ &   1.29   &   0.87   &  30.16   &  11.44  & &   14.68  &  $11.77^{***}$ &  1.24  &   0.59  & 28.12 & 10.66 & & 14.16 &  $11.93^{***}$ &  1.18 &  0.40 & 23.16 &  9.72 \\

\bottomrule

\end{tabular*}
\\[1.0ex]
{\bf Notes:} See Notes of Table 4 for details.
\end{minipage}
\end{center}
\end{table}

\end{landscape}

\newpage
\clearpage

\begin{figure}
\vspace{-1.0cm}
    \caption{Proportion of  correlations declared statistically significant by the FWER  procedures over the trading period when $L=252$ and $N=500$ $(M=124750)$. The results with $\text{BPS}_a$ are omitted due to their close resemblance to those of $\text{BPS}_b$.}
    \label{fig:Wealth}
    \begin{center}        
            \includegraphics[width=1\linewidth]{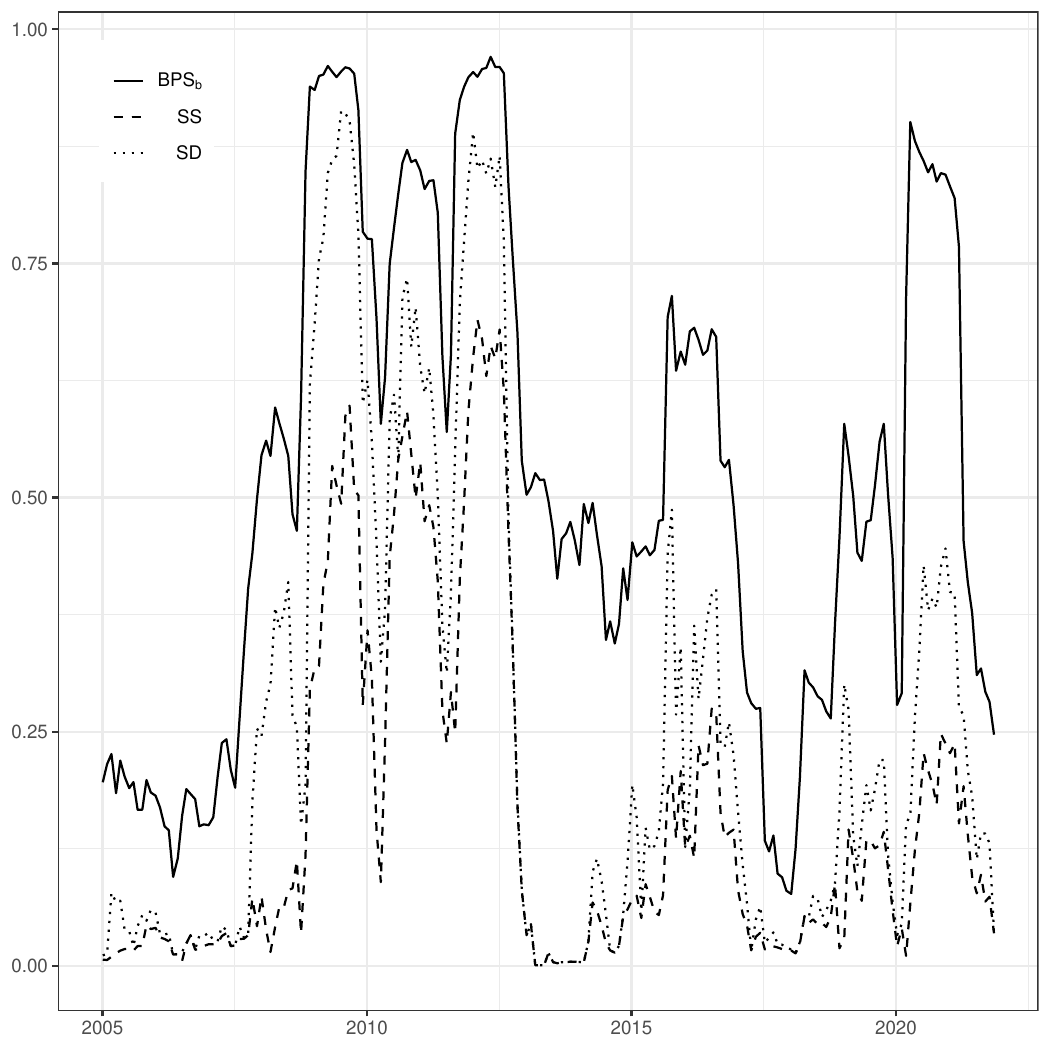}
    \end{center}

\end{figure}

\newpage
\clearpage

\begin{figure}
\vspace{-1.0cm}
    \caption{Proportion of  correlations declared statistically significant by the $k$-FWER  procedures over the trading period when $L=252$ and $N=500$ $(M=124750)$. 
    The results with $\text{SD}_{11}$ and $\text{SD}_{353}$ are omitted due to their close resemblance to those of $\text{SS}_{11}$ and $\text{SS}_{353}$, respectively.
    } 
    \label{fig:Wealth}
    \begin{center}        
            \includegraphics[width=1\linewidth]{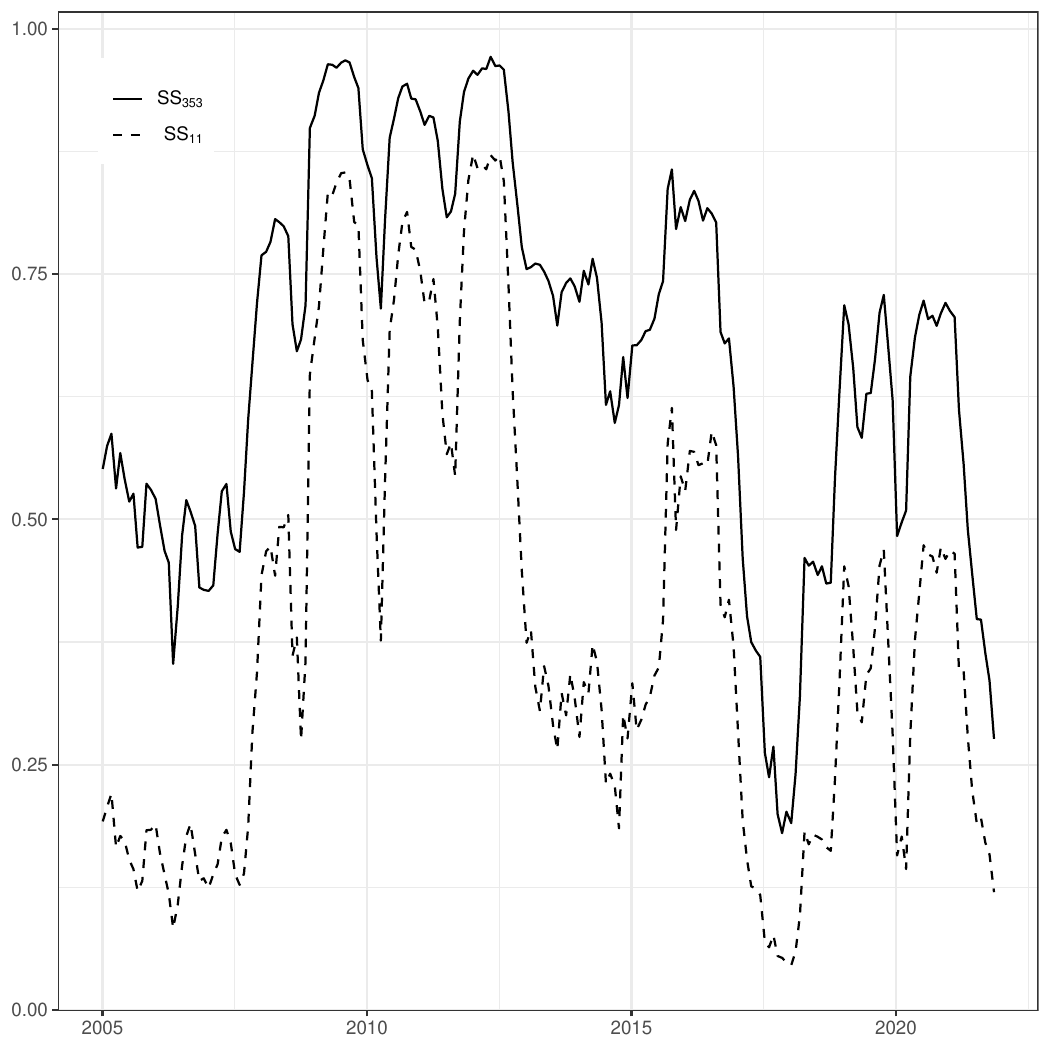}
    \end{center}
        
\end{figure}

\newpage
\clearpage

\begin{figure}
\vspace{-1.0cm}
    \caption{Proportion of  correlations declared statistically significant by the FDP  procedures over the trading period when $L=252$ and $N=500$ $(M=124750)$. The results with $\text{SD}_{0.1}$ are omitted due to their close resemblance to those of $\text{SS}_{0.1}$.
    }
    \label{fig:Wealth}
    \begin{center}        
            \includegraphics[width=1\linewidth]{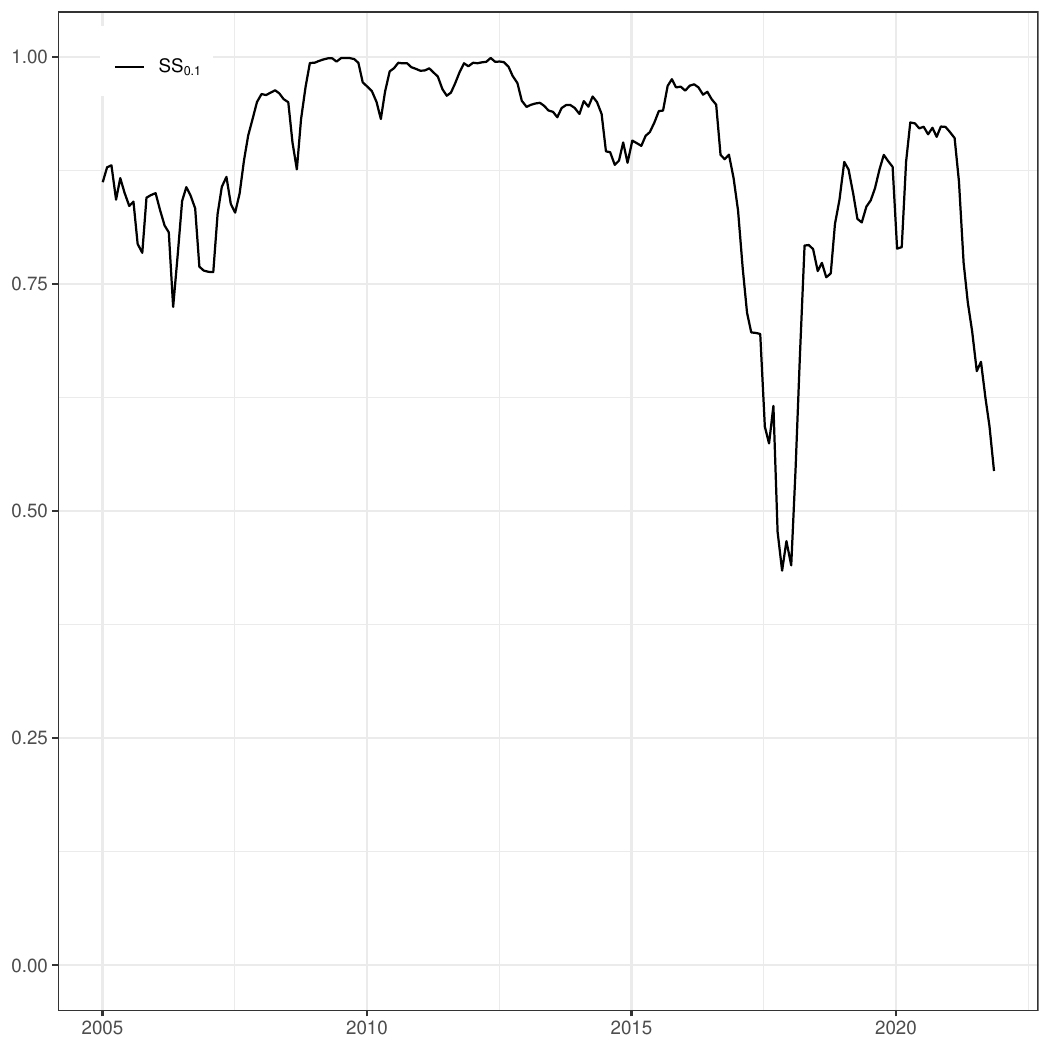}
    \end{center}

\end{figure}

\newpage
\clearpage

\bibliographystyle{chicago}

\bibliography{MT_Regularization8_References}

\newpage

\renewcommand{\baselinestretch}{1.5}
\normalsize

\renewcommand{\thefootnote}{\fnsymbol{footnote}}

\begin{center}

Supplementary material for:

\bigskip

{\textbf{\Large{\sc Regularizing stock return covariance matrices via multiple testing of correlations} } }

\bigskip

\noindent{Richard Luger}\footnotemark[1]  

\smallskip

\noindent{Universit{\'e} Laval, Canada}

\end{center}

\footnotetext[1]{Correspondence to: Department of Finance, Insurance and Real Estate, Laval University, Quebec City, Quebec G1V  0A6, Canada.

\emph{E-mail address:} {richard.luger@fsa.ulaval.ca}. }

\renewcommand{\baselinestretch}{1.5}
\normalsize

\pagenumbering{arabic}

\renewcommand{\thefootnote}{\arabic{footnote}}
\setcounter{footnote}{0}

\section*{Appendix A: Proofs}

\noindent{\bf Proof of Proposition 4.1:} From Theorem 1.3.7 in \citet{Randles-Wolfe:1979} it is known that if $\mathbf Y \stackrel{d}{=} \tilde{\mathbf Y} $ and $\mathcal G(\cdot)$ is a measurable function (possibly matrix-valued) defined on the common support of $\mathbf Y $ and $\tilde{\mathbf Y} $, then $\mathcal G(\mathbf Y) \stackrel{d}{=} \mathcal G(\tilde{\mathbf Y}) $. The fact that the 
$2^{TN}$ possible values (not necessarily distinct) of $\tilde {\boldsymbol \Gamma}$ are equally likely values for $\hat {\boldsymbol \Gamma}$ follows by taking $\mathcal G(\cdot)$ to be the correlation matrix function.  
Observe that  (8) effectively  conditions on $|y_{i,t}|$,  for all $i, t$, since only the signs are randomized.
The zero covariance property in Proposition 4.1 follows from the independence of the Rademacher draws used to generate $\tilde{\mathbf Y}$. \hfill$\square$

\bigskip
\noindent{\bf Proof of Proposition 4.2:}
Proposition 4.1 implies that the pairs \sloppy $(|\tilde \rho_{\ell,1}|, u_1 ),\ldots,(|\tilde \rho_{\ell,B-1}|, u_{B-1} ), (|\hat \rho_{\ell}|, u_B )$
are conditionally \emph{exchangeable} under $H_0$, given $|\mathbf Y|$. 
This means that 
 $\Pr(   R_\text{U}( | \hat \rho_{\ell} | ) = b  \, \vert \, |\mathbf Y|) = 1/B$, for $b=1,\ldots,B$, when $H_0$ holds true, since 
the lexicographic ranks of $B$ exchangeable pairs of random variables are uniformly distributed over  $\{1,\ldots,B\}$.
The stated result then follows on applying Proposition 2.4 in \citet{Dufour:2006} or Lemma 1 in \citet{Romano-Wolf:2005}, and integrating out over $|\mathbf Y|$. \hfill$\square$

\bigskip
\noindent{\bf Proof of Lemma 4.1:} Again this is immediate from Theorem 1.3.7 in \citet{Randles-Wolfe:1979}, this time letting $\mathcal G(\cdot)$ be the composition of the $\text{$k$-max}(\cdot)$, $\text{vechs}(\cdot)$, and correlation matrix functions. Then one simply applies Proposition 2.4 in \citet{Dufour:2006} or Lemma 1 in \citet{Romano-Wolf:2005} upon 
recognition that $(\tilde m^k_{1}, u_1 ),\ldots,(\tilde m^k_{B-1}, u_{B-1} ), (\hat m^k, u_B )$ are conditionally exchangeable under $H_0$, given $|\mathbf Y|$. Integrating out over $|\mathbf Y|$ concludes the proof. \hfill$\square$

\bigskip
\noindent{\bf Proof of Proposition 4.3:}
Let $\text{$k$-min}\big(\tilde p^k_\text{SS}(| \hat \rho_{1} |), \ldots, \tilde p^k_\text{SS}(| \hat \rho_{M} |)  \big)= \tilde p^k_\text{SS}(| \hat \rho_{(k)} |) $ be the 
$k^\text{th}$  smallest value when the $p$-values are ordered as $\tilde p^k_\text{SS}(| \hat \rho_{(1)} |) \le \tilde p^k_\text{SS}(| \hat \rho_{(2)} |) \le \ldots \le \tilde p^k_\text{SS}(| \hat \rho_{(M)} |)  $.  
Rejecting $H_{\ell}$  when $   \tilde p^k_\text{SS}(| \hat \rho_{\ell} |) \le \alpha $ means that
\vspace{-0.75cm}
\[
\begin{split}
 \Pr \left(   \text{Reject at least $k$ hypotheses } H_{\ell} \, \vert \, H_{0} \right)    & = \Pr \left(  \text{At least $k$ $p$-values }   \tilde p^k_\text{SS}(| \hat \rho_{\ell} |) \le \alpha\, \vert \, H_{0}    \right) \\
 \end{split}
\]
\vspace{-1.0cm}
\[
\begin{split}
& = \Pr \bigg(  \text{$k$-min}\big(\tilde p^k_\text{SS}(| \hat \rho_{1} |), \ldots, \tilde p^k_\text{SS}(| \hat \rho_{M} |)  \big) \le \alpha \, \vert \, H_{0} \bigg) \\
& =  \Pr \bigg( \tilde p_\text{U} \big( \text{$k$-max}(|\hat{ \boldsymbol \rho}|) \big)  \le \alpha \, \vert \, H_{0}  \bigg),  
\end{split}
\]
where the last line follows from the equivalence
\[
\bigg\{  \text{$k$-min}\big(\tilde p^k_\text{SS}(| \hat \rho_{1} |), \ldots, \tilde p^k_\text{SS}(| \hat \rho_{M} |)  \big)  \le \alpha  \bigg\}  \Longleftrightarrow \bigg\{  \tilde p_\text{U} \big( \text{$k$-max}(|\hat{ \boldsymbol \rho}|)  \big) \le \alpha  \bigg\},
\]
assuming the same artificial samples $\tilde{\mathbf Y}_b$, $b=1,\ldots,B-1$, and uniform draws $u_b$, $b=1,\ldots,B$, are used in Algorithms 4.1 and 4.2. 
Lemma 4.1 shows that $ \Pr \left(  \tilde p_\text{U} \big( \text{$k$-max}(|\hat{ \boldsymbol \rho}|)  \big) \le \alpha \right) = \alpha$ under $H_0$, which establishes the intended result. \hfill$\square$

\bigskip
\noindent{\bf Proof of Proposition 4.4:}
From the cardinality relationship $|\mathcal M_0| \le |\mathcal M|$, it is straightforward to see that the inequality
\[
 \text{$k$-min}\Bigl( \tilde p^k_\text{SS}\bigl(| \hat \rho_{\ell} | \bigr): \ell \in \mathcal M  \Bigr)  \le  \text{$k$-min}\Bigl( \tilde p^k_\text{SS}\bigl(| \hat \rho_{\ell} | \bigr): \ell \in \mathcal M_0  \Bigr)
\]
always holds. Then, since 
$H_0 \subseteq \bigcap_{\ell \in \mathcal M_0}  H_{\ell}$ (i.e., $\{ H_0 \text{ true} \} \Rightarrow \{ \bigcap_{\ell \in \mathcal M_0}  H_{\ell}  \text{ true} \}$), for every possible choice $\mathcal M_0$, 
it follows  that   
\[
\Pr \Bigl(  \text{$k$-min}\Bigl( \tilde p^k_\text{SS}\bigl(| \hat \rho_{\ell} | \bigr): \ell \in \mathcal M_0  \Bigr) \le \alpha\, \big\vert \, \bigcap_{\ell \in \mathcal M_0}  H_{\ell}  \Bigr) \le \Pr \Bigl( \text{$k$-min}\Bigl( \tilde p^k_\text{SS}\bigl(| \hat \rho_{\ell} | \bigr): \ell \in \mathcal M  \Bigr) \le \alpha\, \big\vert \, H_{0}  \Bigr),
\] 
wherein the left-hand side equals $\Pr \left(   \text{Reject at least $k$ hypotheses } H_{\ell}, \, \ell \in \mathcal M_0 \, \vert \, \bigcap_{\ell \in \mathcal M_0}  H_{\ell}  \right)  $, while the right-hand side equals $\alpha$ by virtue of Proposition 4.3.
Therefore, the probability of $k$ or more  false rejections occurring under the partial null hypothesis $\bigcap_{\ell \in \mathcal M_0}  H_{\ell}$ is bounded from above by the probability of $k$ or more  false rejections occurring when  the more restrictive $H_0$ is imposed. \hfill$\square$

It is interesting to note that the logic of this bound is similar to that of the bounds Monte Carlo test in \citet[][Theorem 4.1]{Dufour-Khalaf:2002}; see also \citet{Dufour:1989}.

\bigskip
\noindent{\bf Proof of Proposition 4.5:} To see first that Algorithm 4.3 has weak control of the $k$-FWER, note the equivalence
\[
\left\{  \text{Reject at least $k$ hypotheses } H_{\ell}  \right\} \Longleftrightarrow   \left\{   \tilde p^k_\text{SD}(| \hat \rho_{\pi_k} |) \le \alpha    \right\}
\]
and the implication
\[
\tilde m^k_k = \text{$k$-max}(|\tilde{ \boldsymbol \rho}|)  \Longrightarrow    \tilde p^k_\text{SD}\big(| \hat \rho_{\pi_k} | \big) =   \tilde p_\text{U}\big( \text{$k$-max}(|\tilde{ \boldsymbol \rho}|)  \big) ,
\]
which holds by construction, conditional on the same underlying $\tilde{\mathbf Y}_b$, $b=1,\ldots,B-1$, and $u_b$, $b=1,\ldots,B$, being used in Algorithms 4.1 and 4.3.
Therefore the SD adjustments from Algorithm 4.3 ensure that
$
\Pr\big(  \text{Reject at least $k$ hypotheses } H_{\ell} \, \vert \, H_0  \big) = \Pr \big(  \tilde p_\text{U}( \text{$k$-max}(|\tilde{ \boldsymbol \rho}|) \le \alpha \, \vert \, H_0  \big) = \alpha, 
$
where the last equality holds by virtue of Lemma 4.1.

Next, observe that the monotonicity-enforced $p$-values (from Step 5 of Algorithm 4.3) ensure the following equality:
\[
 \tilde p^k_\text{SD}\big(| \hat \rho_{\pi_k} | \big) = \text{$k$-min}\big(\tilde p^k_\text{SD}(| \hat \rho_{1} |), \ldots, \tilde p^k_\text{SD}(| \hat \rho_{M} |)  \big), 
\]
wherein the right-hand side satisfies the inequality
\[
 \text{$k$-min}\Bigl( \tilde p^k_\text{SD}\bigl(| \hat \rho_{\ell} | \bigr): \ell \in \mathcal M  \Bigr)  \le  \text{$k$-min}\Bigl( \tilde p^k_\text{SD}\bigl(| \hat \rho_{\ell} | \bigr): \ell \in \mathcal M_0  \Bigr),
\]
since  $|\mathcal M_0| \le |\mathcal M|$. 
As seen before $H_0 \subseteq \bigcap_{\ell \in \mathcal M_0}  H_{\ell}$, for every possible choice $\mathcal M_0$, which here implies the following probability inequality: 
\[
\Pr \Bigl(  \text{$k$-min}\Bigl( \tilde p^k_\text{SD}\bigl(| \hat \rho_{\ell} | \bigr): \ell \in \mathcal M_0  \Bigr) \le \alpha\, \big\vert \, \bigcap_{\ell \in \mathcal M_0}  H_{\ell}  \Bigr) \le \Pr \Bigl( \text{$k$-min}\Bigl( \tilde p^k_\text{SD}\bigl(| \hat \rho_{\ell} | \bigr): \ell \in \mathcal M  \Bigr) \le \alpha\, \big\vert \, H_{0}  \Bigr),
\] 
where the left-hand side equals $\Pr \left(   \text{Reject at least $k$ hypotheses } H_{\ell}, \, \ell \in \mathcal M_0 \, \vert \, \bigcap_{\ell \in \mathcal M_0}  H_{\ell}  \right)  $, while for the right-hand side it is the case that
\[
\Pr \Bigl( \text{$k$-min}\Bigl( \tilde p^k_\text{SD}\bigl(| \hat \rho_{\ell} | \bigr): \ell \in \mathcal M  \Bigr) \le \alpha\, \big\vert \, H_{0}  \Bigr) = 
\Pr \Bigl( \tilde p^k_\text{SD}\big(| \hat \rho_{\pi_k} | \big) \le \alpha\, \big\vert \, H_{0}  \Bigr) = \alpha,
\]
by virtue of weak control.
Once again, the probability of $k$ or more  false rejections occurring under $\bigcap_{\ell \in \mathcal M_0}  H_{\ell}$ is bounded from above by the probability of 
$k$ or more  false rejections occurring under the restrictions of the global intersection null hypothesis, $H_0$. \hfill$\square$

\bigskip

\noindent{\bf Proof of Proposition 4.6:} The use of a $k$-FWER-controlling procedure at each step of  Algorithm 4.4 ensures that $\Pr(F_{k^{\ast}} \ge k^{\ast}   \, \vert \, \bigcap_{\ell \in \mathcal M_0}  H_{\ell} ) \le \alpha$, which implies  $\Pr( F_{k^{\ast}} \ge \gamma(R_{k^{\ast}} + 1)  \, \vert \, \bigcap_{\ell \in \mathcal M_0}  H_{\ell} ) \le \alpha$ since $k^{\ast} \le \gamma(R_{k^{\ast}} + 1)$ by construction.
This last probability statement is equivalent to 
 $\Pr( F_{k^{\ast}}/(R_{k^{\ast}} + 1) \ge \gamma   \, \vert \, \bigcap_{\ell \in \mathcal M_0}  H_{\ell} ) \le \alpha$, which in turn implies that 
$\Pr( F_{k^{\ast}}/R_{k^{\ast}} > \gamma  \, \vert \, \bigcap_{\ell \in \mathcal M_0}  H_{\ell} ) \le \alpha$.
From the definition of $\text{$k$-max}(\cdot)$ it is easy to see that the $k$-FWER-controlling procedure 
is a monotone increasing function of $k$, meaning that $k_1 < k_2 \Rightarrow R_1 \le R_2$.\footnote{In fact, any hypothesis rejected by the $k_1$-FWER procedure is also rejected by the $k_2$-FWER procedure.} The value $k^*$ is thus the largest value of $k$ such that $k \le \gamma(R_k + 1)$, giving the tightest possible bound. \hfill$\square$

\section*{Appendix B: Additional numerical results}

Table B1 reports the average Frobenius norm of the matrix losses $\Delta(\hat {\boldsymbol \Sigma}, {\boldsymbol \Sigma}) = \hat {\boldsymbol \Sigma} - {\boldsymbol \Sigma}$ for the covariance matrix estimators obtained from the FWER, $k$-FWER, and FDP test procedures.
To further the comparisons, the table also reports the Frobenius norm  losses for the sample covariance matrix (Sample) and two shrinkage covariance matrix estimators. The latter are based on 
the linear shrinkage (LS) method of \citet{Ledoit-Wolf:2004} that shrinks the sample covariance matrix towards the identity matrix, and the second one uses  the 
non-linear shrinkage (NLS) estimator proposed by  \citet{Ledoit-Wolf:2015}.\footnote{Specifically, the Ledoit-Wolf shrinkage covariance matrix estimates are computed with the {\tt linshrink\_cov} and {\tt nlshrink\_cov} commands available with the R  package `nlshrink' \citep{Ramprasad:2016}.} The LS and NLS methods are seen to perform particularly well as $N$ increases.

\begin{table}
\begin{center}
\begin{minipage}{6.7in}
\footnotesize
{\bf Table B1.} Frobenius norm losses of regularized covariance matrix estimators when $\delta=0.9$
\\[0.5ex]
\begin{tabular*}{\textwidth}{@{\extracolsep{\fill}}  lrrrrrrrrrrrr}
\toprule

				                 &     		&	  \multicolumn{3}{c}{Normal}	    & 	&   \multicolumn{3}{c}{ $t_{12}$ } 	 & 	&   \multicolumn{3}{c}{ $t_{6}$ } \\

         \cline{3-5} \cline{7-9} \cline{11-13} \\[-2.0ex]
  
                    			&  $T=$     &   63      &   126     &   252         &   &   63      &   126     &  252       &   &   63    &   126        &  252    \\

\midrule
\multicolumn{13}{l}{Panel A:  $N=25$ $(M=300)$} \\

Sample 			             &          &     1.0       &    0.8    &   0.6          &  &    1.2    &   0.9     &   0.7      &  &   1.4    &    1.2      &  0.9   \\
LS              			 &          &     1.0       &    0.8    &   0.6          &  &    1.2    &   0.9     &   0.7      &  &   1.3    &    1.2      &  0.9   \\
NLS                 		 &          &     1.0       &    0.8    &   0.6          &  &    1.1    &   0.9     &   0.7      &  &   1.4    &    1.2      &  0.9     \\
 $\text{BPS}_a/\text{BPS}_b$ & 			&	  1.4		&	 1.0	&	0.7 	     &	&	 1.5 	&	1.2 	&	0.8      &	&	1.7    &	1.5      &  1.1	\\
 $\text{SS}$	     		 & 			&	  1.7 		&	 1.4	&	1.0 	     &	&	 1.7 	&	1.5 	&	1.2      & 	&	1.9    &	1.7      &  1.4   \\
 $\text{SD}$	     		 & 			&	  1.5 		&	 1.1	&	0.7 	     &	&	 1.6 	&	1.3 	&	0.9      &	&	1.8    &	1.5      &  1.1     \\
 $\text{SS}_{5}$     		 & 			&	  1.4 		&	 1.0	&	0.7 	     &	&	 1.5 	&	1.2 	&	0.8      & 	&	1.7    &	1.4      &  1.0   \\
 $\text{SD}_{5}$     		 & 			&	  1.4 		&	 1.0	&	0.7 	     &	&	 1.5 	&	1.1 	&	0.8      &	&	1.6    &	1.4      &  1.0     \\
 $\text{SS}_{17}$     		 & 			&	  1.2 		&	 0.9	&	0.6 	     &	&	 1.3 	&	1.0 	&	0.7      & 	&	1.5    &	1.2      &  0.9   \\
 $\text{SD}_{17}$     		 & 			&	  1.2 		&	 0.9	&	0.6 	     &	&	 1.3 	&	1.0 	&	0.7      &	&	1.5    &	1.2      &  0.9      \\
 $\text{SS}_{0.1}$     		 & 			&	  1.2 		&	 0.9	&	0.6 	     &	&	 1.3 	&	1.0 	&	0.7      & 	&	1.5    &	1.2      &  0.9   \\
 $\text{SD}_{0.1}$     		 & 			&	  1.2 		&	 0.9	&	0.6 	     &	&	 1.3 	&	1.0 	&	0.7      &	&	1.5    &	1.2      &  0.9     \\

\midrule
\multicolumn{13}{l}{Panel B:  $N=100$ $(M=4950)$} \\

Sample 			             &          &     4.1       &    3.2    &   2.4          &  &    4.8    &   3.7     &   2.8      &  &   5.4    &    4.4      &  3.4   \\
LS              			 &          &     4.0       &    3.1    &   2.4          &  &    4.6    &   3.7     &   2.8      &  &   5.2    &    4.3      &  3.3   \\
NLS                 		 &          &     3.8       &    3.0    &   2.3          &  &    4.5    &   3.6     &   2.7      &  &   5.2    &    4.3      &  3.3     \\
 $\text{BPS}_a/\text{BPS}_b$ & 			&	  7.9		&	 7.4	&	6.5 	     &	&	 8.0 	&	7.5 	&	6.9      &	&	8.3    &	7.9      &  7.5	\\
 $\text{SS}$	     		 & 			&	  8.4 		&	 8.1 	&	7.7 	     &	&	 8.4 	&	8.1 	&	7.8      & 	&	8.6    &	8.3      &  8.0   \\
 $\text{SD}$	     		 & 			&	  8.1 		&	 7.6	&	6.8 	     &	&	 8.2 	&	7.7 	&	7.1      &	&	8.4    &	8.0      &  7.6     \\
 $\text{SS}_{8}$     		 & 			&	  8.0 		&	 7.5	&	6.9 	     &	&	 8.0 	&	7.6 	&	7.1      & 	&	8.2    &	7.9      &  7.4   \\
 $\text{SD}_{8}$     		 & 			&	  7.9 		&	 7.4	&	6.7 	     &	&	 8.0 	&	7.6 	&	6.9      &	&	8.2    &	7.8      &  7.3     \\
 $\text{SS}_{70}/\text{SD}_{70}$ &		&	  7.4 		&	 6.6	&	5.4 	     &	&	 7.5 	&	6.8 	&	5.9      &	&	7.8    &	7.3      &  6.5     \\
 $\text{SS}_{0.1}/\text{SD}_{0.1}$ &	&	  6.6 		&	 5.2	&	3.4 	     &	&	 6.8 	&	5.6 	&	4.0      &	&	7.2    &	6.3      &  5.0     \\

\midrule
\multicolumn{13}{l}{Panel C:  $N=500$ $(M=124750)$} \\

Sample 			             &          &     19.9      &    15.6   &   11.5         &  &    22.8   &   17.8    &   13.1     &  &   27.0    &   21.7     &  17.1   \\
LS              			 &          &     19.6      &    15.5   &   11.5         &  &    22.1   &   17.5    &   13.1     &  &   25.6    &   20.8     &  16.8   \\
NLS                 		 &          &     18.5      &    14.8   &   10.9         &  &    21.2   &   16.9    &   12.5     &  &   25.6    &   20.8     &  16.6     \\
 $\text{BPS}_a/\text{BPS}_b$ & 			&	  44.0		&	 43.5	&	42.7	     &	&	 44.1 	&	43.7 	&	43.1     &	&	44.4	&	44.1     &  43.8	\\
 $\text{SS}$	     		 & 			&	  44.4		&	 44.2	&	43.8 	     &	&	 44.5 	&	44.2 	&	43.9     & 	&	44.6  	&	44.3     &  44.1   \\
 $\text{SD}$	     		 & 			&	  44.2		&	 43.8	&	43.0 	     &	&	 44.3  	&	43.9 	&	43.3     &	&	44.4    &	44.1     &  43.8     \\
 $\text{SS}_{11}$     		 & 			&	  44.2   	&	 43.8	&	43.3 	     &	&	 44.2 	&	43.9 	&	43.4     & 	&	44.3    &	44.1     &  43.7    \\
 $\text{SD}_{11}$     		 & 			&	  44.1		&	 43.7	&	43.0 	     &	&	 44.2 	&	43.8 	&	43.3     &	&	44.3    &	44.0     &  43.6     \\
 $\text{SS}_{353}/\text{SD}_{353}$ &	&	  43.6		&	 42.9	&	41.9 	     &	&	 43.7 	&	43.1 	&	42.2     &	&	43.9    &	43.4     &  42.7     \\
 $\text{SS}_{0.1}/\text{SD}_{0.1}$ &	&	  41.6		&	 39.4	&	35.6 	     &	&	 42.0 	&	40.1 	&	36.9     &	&	42.5    &	41.1     &  38.8     \\

  \bottomrule
\end{tabular*}
\\[1.0ex]
Notes:  The results for the $\text{BPS}$ procedures are based on strong FWER-adjusted critical values. The  results for the $k$-FWER procedures, 
 $\text{SS}_{k}$ and $\text{SD}_{k}$, use $k=\lfloor \log{M} \rfloor$ and $k=\lfloor \sqrt{M} \rfloor$; while those for the FDP procedures, 
$\text{SS}_{\gamma}$ and $\text{SD}_{\gamma}$ use $\gamma=0.1.$
When the value of $M$ is large, the $\text{SS}$ and $\text{SD}$ results converge, and in those cases the table presents the results on the same line.

\end{minipage}
\end{center}
\end{table}

\newpage
\clearpage

\thispagestyle{empty}

\begin{landscape}

\begin{table}
\begin{center}
\hspace*{-25pt}
\begin{minipage}{9.85in}
\footnotesize
{\bf Table B2.} Out-of-sample performance of the portfolio strategies when $N=25$ $(M=300)$ 
\\[0.5ex]
\begin{tabular*}{\textwidth}{@{\extracolsep{\fill}}  lcllllllcllllllcllllll}
\toprule

                   						& &      \multicolumn{6}{c}{$L=63$}                           & & \multicolumn{6}{c}{$L=126$}                            & &   \multicolumn{6}{c}{$L=252$}     \\

                                            \cline{3-8} \cline{10-15} \cline{17-22} \\[-2.0ex]
  
                   						         & &   AV      &   SD          &   IR   &   TO &  MDD     &  TW     & &  AV    &    SD          &  IR    &  TO   & MDD   &  TW   & &  AV   &  SD           & IR    &  TO   & MDD    &  TW  \\
\midrule 
  \multicolumn{22}{l}{Panel A: Short selling allowed} \\
 
Sample            					           & &   8.60   &  $16.20^{***}$ &  0.53  & 3.31 &  36.25   &  3.42   & &  12.01 &  $14.63^{**}$  &  0.82  &  1.66  & 30.07 & 6.36 & & 12.58 & $13.62^{***}$ & 0.92  &  0.92 &  26.85  & 7.18  \\
EW                					           & &  14.91   &  $18.43^{***}$ &  0.80  & 0.11 &  42.19   &  9.34   & &  15.06 &  $18.49^{***}$ &  0.81  &  0.11  & 41.04 & 9.57 & & 15.09 & $18.50^{***}$ & 0.81  &  0.10 &  40.80  & 9.61  \\
VT                					           & &  13.32   &  $15.95^{**}$  &  0.83  & 0.22 &  33.71   &  7.67   & &  13.25 &  $16.04$   	  &  0.82  &  0.15  & 34.11 & 7.56 & & 13.31 & $16.13^{**}$  & 0.82  &  0.12 &  33.53  & 7.63  \\
$\text{BPS}_a$    					           & &  14.13   &  $15.74^{**}$  &  0.89  & 1.24 &  30.42   &  8.86   & &  11.43 &  $14.98^{*}$   &  0.76  &  1.29  & 30.68 & 5.72 & & 15.29 & $15.20^{*}$   & 1.00  &  1.38 &  34.88  & 10.93 \\
$\text{BPS}_b$    					           & &  14.19   &  $\bf 15.29$   &  0.92  & 1.36 &  29.16   &  9.05   & &  12.39 &  $\bf 15.47$   &  0.80  &  1.36  & 31.56 & 6.64 & & 13.80 & $\bf 14.95$   & 0.92  &  1.36 &  34.77  & 8.54  \\
SS                					           & &  13.30   &  $15.81^{*}$   &  0.84  & 0.96 &  40.17   &  7.68   & &  12.42 &  $15.65$   	  &  0.79  &  0.96  & 31.47 & 6.65 & & 12.45 & $15.59$       & 0.79  &  1.01 &  29.68  & 6.68  \\
SD                					           & &  14.60   &  $15.45$   	 &  0.94  & 1.22 &  24.91   &  9.67   & &  12.54 &  $15.47$   	  &  0.81  &  1.32  & 28.50 & 6.82 & & 13.28 & $14.78$       & 0.89  &  1.30 &  34.42  & 7.85  \\
$\text{SS}_{5}$ 					           & &  13.69   &  $15.28$   	 &  0.89  & 1.37 &  27.23   &  8.32   & &  14.04 &  $15.09$   	  &  0.93  &  1.28  & 27.47 & 8.87 & & 12.63 & $15.46$       & 0.81  &  1.41 &  28.88  & 6.92  \\
$\text{SD}_{5}$  							   & &  14.44   &  $15.24$   	 &  0.94  & 1.44 &  24.03   &  9.46   & &  12.72 &  $15.30$   	  &  0.83  &  1.38  & 30.47 & 7.06 & & 15.28 & $15.02$       & 1.01  &  1.37 &  32.38  & 10.96 \\
$\text{SS}_{17}$ 							   & &  13.42   &  $14.79^{**}$  &  0.90  & 1.61 &  26.92   &  8.05   & &  12.39 &  $15.17$   	  &  0.81  &  1.55  & 29.50 & 6.69 & & 14.29 & $15.44$       & 0.92  &  1.38 &  29.60  & 9.17 \\
$\text{SD}_{17}$ 							   & &  14.01   &  $15.39$   	 &  0.91  & 1.71 &  27.65   &  8.76   & &  13.96 &  $15.27$   	  &  0.91  &  1.53  & 31.31 & 8.70 & & 14.22 & $14.93$       & 0.95  &  1.38 &  31.52  & 9.17 \\
$\text{SS}_{0.1}$			                   & &  14.39   &  $15.38$   	 &  0.93  & 1.69 &  32.40   &  9.34   & &  12.43 &  $15.80^{*}$   &  0.78  &  1.66  & 31.02 & 6.63 & & 12.93 & $15.20$       & 0.85  &  1.33 &  31.18  & 7.33 \\
$\text{SD}_{0.1}$             		           & &  13.13   &  $15.17$   	 &  0.86  & 1.75 &  33.37   &  7.58   & &  13.42 &  $15.37$   	  &  0.87  &  1.60  & 32.48 & 7.93 & & 14.51 & $14.62^{*}$   & 0.99  &  1.26 &  30.56  & 9.72 \\
LS                					           & &  12.66   &  $13.72^{***}$ &  0.92  & 1.12 &  26.24   &  7.25   & &  12.51 &  $13.70^{***}$ &  0.91  &  0.82  & 29.17 & 7.08 & & 12.57 & $13.45^{***}$ & 0.93  &  0.58 &  25.71  & 7.20  \\
NLS               					           & &  11.73   &  $13.71^{***}$ &  0.85  & 1.77 &  25.17   &  6.20   & &  11.71 &  $13.77^{***}$ &  0.85  &  1.24  & 30.00 & 6.18 & & 12.20 & $13.39^{***}$ & 0.91  &  0.82 &  26.10  & 6.77  \\[1.0ex]

  \multicolumn{22}{l}{Panel B: Short selling prohibited} \\

Sample             						         & &  12.32  &  $13.97^{***}$  &  0.88   & 0.71 &  29.41   &  6.81   & &  12.10 &   $13.92^{***}$ & 0.86 &  0.45  & 29.28 & 6.57 & & 11.45 & $13.74^{***}$ & 0.83  &  0.30 & 28.19  & 5.92 \\
EW                 						         & &  14.91  &  $18.43^{***}$  &  0.80   & 0.11 &  42.19   &  9.34   & &  15.06 &   $18.49^{***}$ & 0.81 &  0.11  & 41.04 & 9.57 & & 15.09 & $18.50^{***}$ & 0.81  &  0.10 & 40.80  & 9.61 \\
VT                 						         & &  13.32  &  $15.95^{***}$  &  0.83   & 0.22 &  33.71   &  7.67   & &  13.25 &   $16.04^{***}$ & 0.82 &  0.15  & 34.11 & 7.56 & & 13.31 & $16.13^{***}$ & 0.82  &  0.12 & 33.53  & 7.63 \\
$\text{BPS}_a$     						         & &  13.65  &  $14.63^{***}$  &  0.93   & 0.55 &  28.93   &  8.40   & &  12.59 &   $14.36$       & 0.87 &  0.48  & 30.46 & 7.06 & & 12.56 & $14.09$       & 0.89  &  0.39 & 29.89  & 7.08 \\
$\text{BPS}_b$     						         & &  13.68  &  $\bf 14.53$    &  0.94   & 0.57 &  29.29   &  8.47   & &  12.89 &   $\bf 14.35$   & 0.89 &  0.48  & 30.47 & 7.44 & & 12.12 & $\bf 14.02$   & 0.86  &  0.39 & 29.63  & 6.58 \\
SS                 						         & &  13.74  &  $15.09^{***}$  &  0.91   & 0.46 &  29.58   &  8.42   & &  12.96 &   $14.97^{***}$ & 0.86 &  0.41  & 30.85 & 7.40 & & 12.30 & $14.64^{***}$ & 0.83  &  0.37 & 30.45  & 6.67 \\
SD                 						         & &  14.10  &  $14.70^{***}$  &  0.95   & 0.54 &  28.20   &  9.05   & &  12.21 &   $14.48$       & 0.84 &  0.49  & 29.76 & 6.60 & & 12.40 & $14.01$       & 0.88  &  0.39 & 29.55  & 6.90 \\
$\text{SS}_{5}$ 								 & &  13.70  &  $14.57$        &  0.94   & 0.59 &  29.10   &  8.48   & &  12.58 &   $14.40$       & 0.87 &  0.48  & 30.05 & 7.04 & & 12.75 & $14.24^{*}$   & 0.89  &  0.39 & 28.47  & 7.27 \\
$\text{SD}_{5}$  								 & &  14.01  &  $14.49$        &  0.96   & 0.61 &  28.69   &  8.95   & &  12.33 &   $14.30$       & 0.86 &  0.49  & 29.43 & 6.77 & & 12.58 & $13.92^{*}$   & 0.90  &  0.37 & 28.95  & 7.13 \\
$\text{SS}_{17}$ 								 & &  13.45  &  $14.29^{***}$  &  0.94   & 0.65 &  29.16   &  8.19   & &  12.73 &   $14.14^{***}$ & 0.90 &  0.51  & 28.71 & 7.27 & & 12.43 & $13.81^{***}$ & 0.90  &  0.37 & 28.45  & 6.97 \\
$\text{SD}_{17}$ 							     & &  13.65  &  $14.31^{***}$  &  0.95   & 0.65 &  29.17   &  8.47   & &  13.09 &   $14.20^{***}$ & 0.92 &  0.50  & 29.00 & 7.72 & & 12.26 & $13.94^{**}$  & 0.87  &  0.37 & 29.25  & 6.75 \\
$\text{SS}_{0.1}$             			         & &  13.85  &  $14.32^{***}$  &  0.96   & 0.65 &  28.68   &  8.75   & &  12.92 &   $14.12^{***}$ & 0.91 &  0.50  & 29.00 & 7.52 & & 12.15 & $13.82^{***}$ & 0.87  &  0.35 & 29.35  & 6.65 \\
$\text{SD}_{0.1}$			                     & &  13.65  &  $14.28^{***}$  &  0.95   & 0.65 &  28.46   &  8.46   & &  12.70 &   $14.10^{***}$ & 0.90 &  0.51  & 28.85 & 7.25 & & 12.11 & $13.91^{***}$ & 0.87  &  0.35 & 29.01  & 6.58  \\
LS                 						         & &  13.21  &  $13.95^{***}$  &  0.94   & 0.58 &  30.22   &  7.92   & &  12.69 &   $13.86^{***}$ & 0.91 &  0.40  & 30.39 & 7.27 & & 11.70 & $13.69^{***}$ & 0.85  &  0.28 & 27.76  & 6.17 \\
NLS                						         & &  12.99  &  $13.89^{***}$  &  0.93   & 0.59 &  30.63   &  7.65   & &  12.61 &   $13.86^{***}$ & 0.91 &  0.41  & 30.30 & 7.18 & & 11.54 & $13.68^{***}$ & 0.84  &  0.29 & 28.01  & 6.01 \\

\bottomrule

\end{tabular*}
\\[1.0ex]
{\bf Notes:} This table reports the  
annualized mean (AV, in \%), 
annualized standard deviation (SD, in \%), 
 information ratio (IR),
 average turnover (TO),
maximum drawdown (MDD, in \%),
and terminal wealth (TW, in \$)
  achieved by the rolling-window  portfolio strategies with monthly rebalancing, given an initial wealth of \$1. 
The trading period is from January 3, 2005 to  December 31, 2021, and
$L$ is the length (in days) of the estimation window. 
One, two, and three stars indicate that SD is statistically different from  SD of the benchmark $\text{BPS}_b$ strategy (in bold)
at the 10\%, 5\%, and 1\% levels, respectively.

\end{minipage}
\hspace{0.5in} 
\end{center}
\end{table}

\end{landscape}

\newpage
\clearpage

\section*{Appendix C: Computation times}

\smallskip

\begin{table}[ht]
\begin{center}
\begin{minipage}{6.7in}
\vspace{-0.6cm}
\footnotesize
{\bf Table C1.} \\[0.5ex]
\begin{tabular*}{\textwidth}{@{\extracolsep{\fill}}  lrrrrrrrr}
\toprule

			            					&      &	  \multicolumn{3}{c}{$N=100$ $(M=4950)$}	& 	&   \multicolumn{3}{c}{$N=500$ $(M=124750)$}  \\

         										\cline{3-5} \cline{7-9}  \\[-2.0ex]
  
			            					&	$T=$	& 	63	  &	126       &	252		&	&  	63		&	126	      &	252	  			\\

\midrule

 $\text{SS}$	     		 				& 		  &	  0.2 secs	&	0.3 secs	&	0.4 secs 	     &	&    9.9 secs 	&	11.9 secs 	&	15.5 secs   \\
 $\text{SD}$	     						& 	   	  &	  0.2 secs  &	0.3 secs	&	0.4 secs 	     &	&	10.7 secs 	&	13.0 secs 	&	16.7 secs   \\
$\text{SS}_{\lfloor \log{M} \rfloor}$  		&         &   0.2 secs  &   0.3 secs    &   0.4 secs     	 &  &    9.9 secs   &   12.0 secs   &   15.4 secs       \\
$\text{SD}_{\lfloor \log{M} \rfloor}$  		&         &   0.2 secs  &   0.3 secs    &   0.4 secs     	 &  &   10.9 secs   &   12.9 secs   &   16.4 secs      \\
$\text{SS}_{\lfloor \sqrt{M} \rfloor}$ 		&         &   0.2 secs  &   0.3 secs    &   0.4 secs     	 &  &   10.0 secs   &   12.0 secs   &   15.7 secs      \\
$\text{SD}_{\lfloor \sqrt{M} \rfloor}$ 		&         &   0.2 secs  &   0.3 secs    &   0.4 secs         &  &   10.8 secs   &   12.9 secs   &   16.8 secs     \\[1.0ex]
 $\text{SS}_{0.1}$     		 				& 	  	  &	  2.8 secs 	&	3.5 secs	&	4.8 secs 	     &	&	 3.1 mins 	&	 3.7 mins 	&	 4.9 mins  \\
 $\text{SD}_{0.1}$     		 				& 		  &	  3.6 secs	&	4.4 secs	&	5.9 secs 	     &	&	 3.7 mins 	&	 4.4 mins 	&	 5.8 mins  \\

\bottomrule

\end{tabular*}

\end{minipage}
\end{center}
\end{table}

\vspace{-0.6cm}
This table presents the computing times for the multiplicity-adjusted $p$-values when $B=100.$ The omission of computer hardware details is intentional to emphasize the relative performance as $N$ increases from $100$ to $500$. 
Computing times increase markedly, especially for the FDP procedures, even though the bisection method described in Algorithm 4.5 is used to speed up the computations.
For instance, with $T=252,$ the $\text{SD}_{0.1}$ procedure takes about 6 seconds when $N=100$ but almost 6 minutes when $N=500$, which represents  a sixty-fold increase.

\end{document}